\documentclass[12pt]{article}
\usepackage{amssymb,amsmath,latexsym,graphicx, bm, mathrsfs, hyperref}
\usepackage{graphics,graphicx, cancel}
\usepackage[dvipsnames]{xcolor}

\usepackage[british]{datetime2}

\begin{document}

\title{More on Jacobi metric: \\ 
Randers-Finsler metrics, frame dragging and \\ 
geometrisation techniques 
\vspace{-0.25cm} 
\author{Sumanto Chanda \\ \vspace{-0.5cm} \\ 
\textit{Indian Institute of Astrophysics} \\ 
\textit{Block 2, 100 Feet Road, Koramangala,} 
\textit{Bengaluru 560034, India.} \\ 
\texttt{\small sumanto.chanda@iiap.res.in}}}

\maketitle

\thispagestyle{empty}
\vspace{-1.0cm}
\abstract{
In this article, I demonstrate a new method to derive Jacobi 
metrics from Randers-Finsler metrics by introducing a more 
generalised approach to Hamiltonian mechanics for such 
spacetimes and discuss the related applications and properties. 
I introduce Hamiltonian mechanics with the constraint 
for relativistic momentum, including a modification for 
null curves and two applications as exercises: 
derivation of a relativistic harmonic oscillator, and 
analysis of Schwarzschild Randers-Finsler metric. 
Then I describe the main application for constraint 
mechanics in this article: 
a new derivation of Jacobi metric for time-like and 
null curves, comparing the latter with optical metrics. 
After that, I discuss frame dragging with the Jacobi 
metric, and two applications for Randers-Finsler metrics: 
an alternative to Eisenhart lift, and different metrics 
that share the same Jacobi metric. 
}

\vspace{-0.5cm}
\tableofcontents

\numberwithin{equation}{section}

\newpage
\setcounter{page}{1}

\section{Introduction}
\label{sec:intro}

The Jacobi metric describes the projection of a geodesic 
in spacetime onto a hypersurface characterized by its 
energy. 
It is a topic of significant interest that has been studied 
for many years 
\cite{ong,shs,gibbons,cgg1,cgg2,ilgm,cggmw,maraner,bgm,argand,LiJia1,LiJia2}. 
Ong described an interesting application of the theory 
to gravity \cite{ong} to study the curvature of Jacobi 
metric of Newtonian $N$-body problem, while Bera, Ghosh, 
and Majhi used it to study Hawking Radiation \cite{bgm}. 
The basic concept originates in the Maupertuis principle 
from which it is formulated for Hamiltonian systems 
\cite{shs,gibbons,cgg1,cgg2,ilgm,cggmw,maraner}, 
and it has found many applications in gravity 
\cite{ong,bgm,argand,LiJia1,LiJia2}.  
More recently, alongside Gibbons and Guha, I have discussed 
Jacobi metric in the study of geodesic flows \cite{cgg1}, 
an application to the study of Kepler systems \cite{cgg2}, 
and the gravitational magnetoelectric effect for stationary 
spacetimes in \cite{cggmw} alongside Maraner and Werner. 
However, so far it has been discussed only for 
spacetimes described via Lorentzian geometry. 
\smallskip

Lorentzian geometry is a special case of pseudo-Riemannian 
geometry which is used to describe spacetime for the study 
of gravity in general relativity \cite{mtw,carroll} by measuring 
the length of worldline curves on a Lorentzian manifold $M$ 
with the square root of the norm with Lorentzian signature 
on the local tangent space $T_x M$. 
The Jacobi metrics for static Lorentzian metrics are described 
via Riemannian geometry \cite{gibbons,cgg1}. 
However, Jacobi metrics derived from stationary metrics 
depart from the familiar Riemannian form, being described 
as a more general class of metrics known as Randers-Finsler 
(RF) metrics. 
In 1941 Randers \cite{Randers} modified a Riemannian 
metric into a Finsler metric by adding a linear term 
$A = A_i (\bm x) dx^i$, where $A_i (\bm x)$ are potentials. 
This Finsler metric is known as the Randers-Finsler metric, 
which simultaneously accounts for curvature and potential 
functions. 
The Jacobi metrics for stationary Lorentzian metrics are 
described via RF metrics \cite{cggmw}. 
In $n$ dimensions, a RF metric can be written as: 
$$ 
     F_{RF} (x, y) = 
     \sqrt{g_{ij} (\bm x) y^i y^j} + 
     A_i (\bm x) y^i, 
\qquad 
     y = y^i \partial_i \in T_x M, 
\quad 
     i, j = 1, 2, . . . n.
$$  
The equations of motion derived for the above metric 
are best described as Newton's equations of motion 
involving Lorentz force, instead of the geodesic 
equation associated with a Riemannian metric. 
\smallskip

Trivial examples of such RF metrics are described 
in 3 dimensions with Euclidean signature, being 
derived from stationary metrics in 3+1 Lorentzian 
spacetimes either as Jacobi metrics \cite{cggmw}, 
or as optical metrics \cite{ghww} describing 
geodesics of shortest time travel for light-like 
particles. 
However, in recent years there has been a growing 
interest in studying gravity with RF metrics in 3+1 
Lorentzian spacetimes. 
Stavrinos, Basalikos, Triantafyllopoulos and others 
have studied the example of the Schwarzschild RF 
metric \cite{ktbs1,ktbs2,kkst}. 
Heefer and Pfeifer studied gravitational waves and 
extensions of Einstein's gravity in RF spacetime 
\cite{heeffust,heefer,pfewol,pfeifer,hpf}, and Silva 
has described field theory in RF spacetime 
\cite{silva,sma}. 
Considering the current interest in RF metrics in 3+1 
Lorentzian spacetimes, it is necessary to also describe 
a new and more generalised Jacobi metric for such 
spacetimes and the resulting properties, to extend 
the utility of Jacobi metrics to this more generalised 
setting, and explore new relevant applications, one 
of which is the Eisenhart lift. 
\smallskip

In 1929 Eisenhart \cite{eisenhart} compared trajectories 
of dynamical system in classical configuration space in 
$n$ co-ordinates to geodesics in $n+2$-co-ordinates. 
This suggests that they are equivalent, which is evident 
from projection of geodesics into regular Lagrangian systems. 
Based on his work, Gibbons, Duval, Horvathy, Minguzzi 
and many others developed the Eisenhart lift for natural 
Hamiltonian systems 
\cite{lichteich,dbkp,Ettore1,Ettore2,Car,CarGib,cghhz,CarAlv,GalMas,FilGal}, 
which reverses the projection from a regular Lagrangian 
into a geodesic. 
It is a method for geometrising potentials of  non-relativistic 
systems by adding extra degrees of freedom while preserving 
Hamilton's equations of motion, thus converting them into 
relativistic systems, allowing us to use geometric approach 
to study them.
The procedure has found many applications in the study 
of gravity and integrable systems 
\cite{Ettore1,Ettore2,FilGal,dgh,BekMor,CarGal,chr,ForGal,Gal1,cggh,Mor,spds}. 
More recently, the Eisenhart lift was generalised for application 
to field theory \cite{fkp,cg2}. 
\smallskip

In light of the recent growing interest in RF metrics, 
and the popularity of Eisenhart lift as a geometrising 
technique for regular Lagrangian systems, it would 
be interesting to describe similar procedures for RF 
metrics via Jacobi metrics and study the related 
relativistic physics such as the frame-dragging effect. 
In this article, I shall return to the topic of Jacobi metrics 
discussed previously in \cite{cgg1,cggmw},and this time 
generalise its form and derivation for more general RF 
metrics, study its properties such as comparison to optical 
metrics and the frame dragging effect, and introduce 
applications such two new methods of geometrisation 
of potentials such as a new type of Eisenhart lift, and 
different RF metrics sharing the same Jacobi metric. 
In Section 2, I will begin by reviewing preliminaries 
of basic classical mechanics, especially Hamiltonian 
mechanics. 
Here, I will introduce a momentum constraint for RF 
spacetimes that was first introduced in \cite{cggmw}, 
and further developed upon in \cite{cg2}, this time 
including a modification to allow consideration of null 
curves, and demonstrate how Hamilton's equations 
of motion derive from it. 
In Section 3, I will derive the generalised JMRF 
using the constraint and reproduce the Jacobi metric 
for under familiar settings for stationary spacetimes 
discussed in \cite{cgg1,cggmw}. 
Then I will describe the JMRF for null curves and 
compare it to the optical metric. 
In Section 4, I will discuss derivation of frame dragging 
effect using Epstein's Hamiltonian approach \cite{epstein} 
first for stationary Riemannian metrics, then RF metrics, 
and finally for the JMRF metrics. 
Finally, in Section 5, I shall demonstrate two ways 
to geometrise the potentials in the linear additive 
term of the RF metric. 
I will start by showing why the Eisenhart lift faces 
limitations when attempted upon RF metrics, and 
show how to use the JMRF metric to demonstrate 
one way to overcome the problem. 
Then I will discuss how a Riemannian and a RF metric 
can share the same JMRF metric as another method 
of geometrising the linear term potentials, and apply 
the procedure in reverse to the Schwarzschild 
Gullstrand-Painlev\'e metric and study the result 
as an example.

\numberwithin{equation}{section}

\section{Preliminaries: Hamilton's equations of motion for RF spacetimes}
\label{sec:prelim} 

General Relativity \cite{mtw,carroll} can be regarded 
as an advanced version of classical mechanics \cite{goldstein}, 
borrowing and applying many principles from the latter 
to a more sophisticated level in curved space. 
Hamilton's equations of motion are an important 
aspect in the formulation of the Jacobi metric since 
they must be preserved from the original metric, 
and thus their description must be addressed for 
RF metrics. 
In this section, I shall review the associated 
mechanical preliminaries relevant later on in 
this article. 

Although Maupertuis is credited for the principle of least 
action that applies to all physical systems, he had originally 
applied it only to light \cite{maup}, evidence pointing to Euler 
\cite{euler} for intuitively connecting it to mechanics. 
In its modern form, Maupertuis principle has proved very 
relevant in mechanics, providing a linear form of action 
applied in path-integral formulation in quantum mechanics.
\bigskip 

\noindent
If the worldline length $s$ of a curve on $n + 1$ dimensional 
spacetime between two points given by integration of the 
metric $d s$ is parametrised by $\tau$, it can be written in terms 
of a Lagrangian $L$ such that:
\begin{equation} \label{action} 
     s = 
     \int_1^2 d s = 
     \int_1^2 d \tau \ L (\bm x, \bm{\dot x}) 
, \qquad 
     \text{where } \ L = \frac{d s}{d \tau}, 
\quad 
     \bm{\dot x} = \frac{d {\bm x}}{d \tau}.
\end{equation} 
From here on throughout the rest of the article, 
I am declaring that all indices in greek letters represent 
spacetime components such as $\mu, \nu = 0, 1, 2, . . . . $, 
while all indices in english alphabets represent spatial 
components such as $i, j = 1, 2, . . . n$. 
Varying the Lagrangian in \eqref{action} gives us:
\begin{equation} \label{var} 
     \delta L = 
     \left\{ 
          \frac{\partial L}{\partial x^\mu} - 
          \frac{d \ }{d \tau} 
          \left( \frac{\partial L}{\partial \dot{x}^\mu} \right) 
     \right\} 
     \delta x^\mu + 
     \frac{d \ }{d \tau} 
     \left( 
          \frac{\partial L}{\partial \dot{x}^\mu} 
          \delta x^\mu 
     \right).
\end{equation} 
Motion will occur along the curve described by the solution 
of Euler-Lagrange equation, called the geodesic:
\begin{equation} \label{eulag} 
     \frac{\partial L}{\partial x^\mu} - 
     \frac{d p_\mu}{d \tau} = 
     0 
\qquad , \qquad 
     p_\mu = 
     \frac{\partial L}{\partial \dot x^\mu}.
\end{equation} 
The variation of the curve length \eqref{action} close to the 
geodesic \eqref{eulag} is given by the variation at the ends 
of the curve, shown by applying \eqref{eulag} to \eqref{var}:
$$
     \delta s = 
     \int_1^2 d \tau \ 
     \delta L = 
     \left[ 
          p_\mu \; 
          \delta x^\mu 
     \right]_1^2 
\quad \equiv \quad 
     \left[ 
          \frac{\partial s}{\partial x^\mu} 
          \delta x^\mu 
     \right]_1^2 
\qquad \Rightarrow \qquad 
     p_\mu = 
     \frac{\partial s}{\partial x^\mu},
$$ 
from which we can write the Maupertuis principle 
that describes the action along the classical trajectory 
$\bm{x}_{c} (\tau)$ that satisfies the Euler-Lagrange 
equation \eqref{eulag}:
\begin{equation} \label{maup} 
     d s = 
     d \tau \; 
     L_M = 
     \frac{\partial s}{\partial x^\mu} 
     d x^\mu = 
     p_\mu d x^\mu.
\end{equation}
Thus, we can deduce the Maupertuis form of the 
Lagrangian $L_M$ which according to \eqref{maup} 
matches the Lagrangian according to \eqref{action} 
on the classical trajectory $\bm{x}_{c} (\tau)$.
\begin{equation} \label{mauplag} 
     L_M = 
     p_\mu 
     \dot x^\mu_{c} = 
     L (\bm{x}_{c}, \bm{\dot x}_{c}). 
\end{equation} 
from which the overall Hamiltonian $\mathcal H$ 
according to Legendre's principle is a constrained 
at the value of zero: 
\begin{equation} \label{ovham} 
     \mathcal H = 
     p_\mu 
     \dot x^\mu_{c} - 
     L (\bm{x}_{c}, \bm{\dot x}_{c}) 
     = 0. 
\end{equation} 
Usually, when dealing with natural Lagrangian systems 
where the Lagrangian is given by: 
\begin{equation} \label{natlag} 
     L = 
     \frac12 
     g_{\mu \nu} (\bm x) 
     \dot x^\mu 
     \dot x^\nu, 
\end{equation} 
one can see that according to \eqref{ovham} the regular 
Hamiltonian derived via Legendre's principle is given by: 
\begin{equation} \label{regmauplag} 
     p_\mu = 
     \frac{\partial L}{\partial x^\mu} = 
     g_{\mu \nu} (\bm x) 
     \dot x^\nu 
\qquad \Rightarrow \qquad 
     p_\mu 
     \dot x^\mu = 
     g_{\mu \nu} (\bm x) 
     \dot x^\mu 
     \dot x^\nu = 
     2 L, 
\end{equation}  
$$
\Rightarrow \qquad 
     p_\mu 
     \dot x^\mu - 
     L = 
     \frac12 
     g_{\mu \nu} (\bm x) 
     \dot x^\mu 
     \dot x^\nu = 
     \frac12 
     g^{\mu \nu} (\bm x) 
     p_\mu 
     p_\nu, 
$$ 
\begin{equation} \label{regham} 
\Rightarrow \qquad 
     \mathcal H = 
     \frac12 
     g^{\mu \nu} (\bm x) 
     p_\mu 
     p_\nu = 
     0.
\end{equation} 
where we can see that the LHS of \eqref{regham} 
is non-vanishing because the expression of the 
Maupertuis form of the Lagrangian according 
to \eqref{mauplag} does not match Lagrangian 
\eqref{natlag} as seen from \eqref{regmauplag}. 
For Lorentzian metrics, the momenta in 
\eqref{regham} can be non-zero. 
Upon parametrisation wrt time $x^0 = t$, 
the velocity is concealed ($\dot x^0 = \dot t = 1$), 
and the Legendre principle leads to the associated 
momentum $p_0$:
\begin{equation} \label{ham} 
     L = 
     p_i \dot x^i + p_0 
\qquad \Rightarrow \qquad 
     H = 
     - p_0 = p_i \dot x^i - L.
\end{equation}
Since the velocity component for time $t$ is lost 
upon being given the status of a parameter, 
the negative of its conjugate momentum $p_0$ 
defined as the Hamiltonian $H$ is provided 
by the Legendre principle \eqref{ham}. 
Since $H$ is a function of $x$ and $p$, 
the variation of $H$ gives Hamilton's equation 
of motion:
\begin{equation} \label{hameom}
     \frac{d x^i}{d t} = 
     \frac{\partial H}{\partial p_i} 
\qquad , \qquad 
     \frac{d p_i}{d t} = 
     - \frac{\partial H}{\partial x^i}.
\end{equation} 
In case of regular Lagrangian systems, the Lagrangian 
does not match the Maupertuis form \eqref{mauplag}, 
which allows us to formulate the Hamiltonian function 
in the phase space using \eqref{ham}. 
Under the circumstances that the system is independent 
of a co-ordinate $x^0$ referred to as a cyclic co-ordinate, 
the corresponding momentum will be a conserved quantity 
according to \eqref{eulag} and \eqref{hameom}, 
the existence of which is essential for the Eisenhart lift and 
its alternatives for Randers-Finsler metrics.
\begin{equation} \label{consmom} 
     \frac{d p_0}{d \tau} = 
     \frac{\partial L}{\partial x^0} = 
     - \frac{\partial H}{\partial x^0} = 0 
\qquad \Rightarrow \qquad 
     p_0 = q (const). 
\end{equation} 
This results in the associated term in the Maupertuis Lagrangian 
\eqref{mauplag} becoming a total time derivative that can be 
dismissed from the Lagrangian: 
\begin{equation} \label{lageff} 
     L = p_i \dot x^i + q \dot x^0 
\qquad \Rightarrow \qquad 
     L_{eff} = p_i \dot x^i. 
\end{equation} 
Thus, according to \eqref{lageff}, extra degrees of freedom 
can in principle be removed or inserted into the description 
of a particles mechanics. 
It is here that the Hamiltonian proves essential to formulating 
Eisenhart and Jacobi metrics for Hamiltonian systems without 
altering the equations of motion \cite{gibbons,cgg1}. 
However, one has to consider circumstances where a proper 
Hamiltonian cannot be deduced.

In RF metrics \cite{Randers} the first part with the norm 
under the square root accounts for the influence of curvature, 
while the linear term outside accounts for gauge field interaction. 
Sometimes the second part is geometric in origin. 
\begin{equation} \label{rfmetric} 
     ds = 
     \sqrt{
     g_{\mu \nu} (\bm x) 
     d x^\mu 
     d x^\nu 
     } + 
     A_\mu (\bm x) 
     d x^\mu. 
\end{equation} 
Using the Lagrangian $L$ derived from the RF metric 
\eqref{rfmetric} according to \eqref{action}: 
\begin{equation} \label{rflag} 
     L = 
     \frac{d s}{d \tau} = 
     \sqrt{
          g_{\mu \nu} (\bm x) 
          \dot x^\mu 
          \dot x^\nu
     } + 
     A_\mu (\bm x) 
     \dot x^\mu.
\end{equation} 
The canonical momenta $\bm p$ according 
to \eqref{eulag}
\begin{equation} \label{rfcanmom} 
     p_\mu = 
     \frac{\partial L}{\partial \dot x^\mu} = 
     g_{\mu \nu} (\bm x) 
     \frac{d x^\nu}{d \sigma} + 
     A_\mu (\bm x), 
\qquad \text{where } \ 
     d \sigma := 
     \sqrt{g_{\alpha \beta} (\bm x) d x^\alpha d x^\beta}, 
\end{equation} 
leads us to the gauge-covariant momenta 
$\bm \pi$ given by 
\begin{equation} \label{rfgcmom1} 
     \pi_\mu = 
     p_\mu - A_\mu (\bm x) = 
     g_{\mu \nu} (\bm x) 
     \frac{d x^\nu}{d \sigma}. 
\end{equation} 
Here using the canonical momentum \eqref{rfcanmom}, 
we can verify that the expression of the Maupertuis form 
\eqref{mauplag} is exactly identical to the Lagrangian 
for RF metrics \eqref{rflag} 
 $$
     p_\mu 
     \dot x^\mu = 
     \frac{g_{\mu \nu} (\bm x) \dot x^\nu}
     {\sqrt{g_{\alpha \beta} (\bm x) \dot x^\alpha \dot x^\beta}} 
     \dot x^\mu + 
     A_\mu (\bm x) 
     \dot x^\mu = 
     \sqrt{
          g_{\mu \nu} (\bm x) 
          \dot x^\mu 
          \dot x^\nu
     } + 
     A_\mu (\bm x) 
     \dot x^\mu = 
     L. 
$$ 
Thus, in the overall Hamiltonian \eqref{ovham} derived 
according to Legendre's principle given by: 
$$ 
     \mathcal H = 
     p_\mu 
     \dot x^\mu - 
     L = 0, 
$$ 
we can see that the LHS vanishes unlike the regular 
Hamiltonian for natural Lagrangian systems \eqref{regham}, 
which means that the Hamiltonian cannot be formulated 
via Legendre's principle \eqref{ovham}. 
Thus, an alternative generator of equations in phase 
space must be found. 
To this end, I show that the gauge-covariant momenta 
$\bm \pi$ \eqref{rfgcmom1} obey the constraint:
\begin{equation} \label{constraint} 
     \phi (\bm x, \bm p) = 
     \sqrt{
          g^{\mu \nu} (\bm x) 
          \pi_\mu \pi_\nu
     } = 
     \sqrt{
          g_{\mu \nu} (\bm x) 
          \frac{d x^\mu}{d \sigma} 
          \frac{d x^\nu}{d \sigma}} = 1,
\end{equation} 
which acts as a generator of equations of motion \cite{cg2}, 
demonstrated by taking a derivative of the constraint:
\begin{equation} \label{cons2} 
     \frac{d \phi}{d \sigma} = 
     \frac{\partial \phi}{\partial x^\mu} 
     \frac{d x^\mu}{d \sigma} + 
     \frac{\partial \phi}{\partial p_\mu} 
     \frac{d p_\mu}{d \sigma} = 0,
\end{equation}
then I can show by applying \eqref{rfgcmom1} and 
\eqref{constraint} into \eqref{cons2} that one will have:
$$
     \frac{\partial \phi}{\partial p_\mu} = 
     g^{\mu \nu} (\bm x) 
     \pi_\nu = 
     \frac{d x^\mu}{d \sigma}
\qquad \Rightarrow \qquad 
     \frac{\partial \phi}{\partial x^\mu} = 
     - \frac{d p_\mu}{d \sigma}.$$
Thus, we have the constraint equivalent of Hamilton's 
equations of motion:
\begin{equation} \label{conseq} 
     \boxed{
          \frac{d x^\mu}{d \sigma} = 
          \frac{\partial \phi}{\partial p_\mu}, 
     \qquad 
          \frac{d p_\mu}{d \sigma} = 
          - \frac{\partial \phi}{\partial x^\mu}
     }.
\end{equation} 
Under the circumstances that one is dealing with 
spacetime metrics that possess Minkowskian 
signature, there is entirely a possibility that one 
may deal with null curves which lead to the constraint 
becoming undefined according to \eqref{constraint}. 
To overcome this obstacle, we insert an extra auxilary 
co-ordinate $y$ without disturbing the mechanics 
of the system, by re-writing 
the norm $d \sigma$ introduced 
in \eqref{rfcanmom} into: 
\begin{equation} \label{norm} 
     d \sigma := 
     \sqrt{
          \kappa 
          d y^2 + 
          d \Sigma^2 
     }, 
\qquad \text{where } \ 
     d \Sigma^2 := 
     g_{\mu \nu} (\bm x) 
     d x^\mu 
     d x^\nu, 
\end{equation} 
and the RF metric \eqref{rfmetric} 
into: 
\begin{equation} \label{rfmod} 
     d s = 
     \sqrt{
          \kappa 
          d y^2 + 
          d \Sigma^2 
     } + 
     A_\mu (\bm x) 
     d x^\mu, 
\qquad \text{where } \ 
     \kappa = 
     \begin{cases} 
          0 \ , \ d \Sigma^2\neq 0
     \\ 
          1 \ , \ d \Sigma^2 = 0
     \end{cases}. 
\end{equation} 
This way, when we deduce the momenta according 
to \eqref{eulag}, then we can write under the limit 
\eqref{rfmod}: 
\begin{equation} \label{extramom} 
     p_y = 
     \kappa \frac{d y}{d \sigma} = 
     \frac{\kappa d y}{\sqrt{\kappa d y^2 + d \Sigma^2}} = 
     \begin{cases} 
          0 \ , \ d \Sigma^2 \neq 0
     \\ 
          1 \ , \ d \Sigma^2 = 0
     \end{cases}, 
\end{equation} 
the constraint \eqref{constraint} will become: 
\begin{equation} \label{consmod} 
     \phi (\bm{x, p}) = 
     \sqrt{
          p_y^2 + 
          g^{\mu \nu} (\bm x) 
          \pi_\mu 
          \pi_\nu 
     } = 1, 
\qquad \text{where } \ 
     p_y = 
     \begin{cases} 
          0 \ , \ d \sigma^2 \neq 0
     \\ 
          1 \ , \ d \sigma^2 = 0
     \end{cases}. 
\end{equation} 
Although the RF Lagrangian exactly matching 
the Maupertuis form prevents us from deducing 
a Hamiltonian function, it does allow us to determine 
the metric using the Maupertuis principle. 
When starting from the constraint, this is done 
by applying \eqref{rfgcmom1} and the first equation 
of \eqref{conseq} to Maupertuis principle \eqref{maup}:
\begin{equation} \label{maup2} 
     d s = 
     p_\mu d x^\mu = 
     g_{\mu \nu} (\bm x) 
     \frac{\partial \phi}{\partial p_\mu} 
     d x^\mu + 
     A_\mu (\bm x) d x^\mu = 
     d \sigma + 
     A_\mu (\bm x) d x^\mu.
\end{equation}
From \eqref{constraint}, we can write 
$d \sigma = 
\sqrt{g_{\mu \nu} (\bm x) d x^\mu d x^\nu}$,
which upon application to \eqref{maup2}, will give 
us the original RF metric \eqref{rfmetric}. 
Thus, we can see that mechanics with the momentum 
constraint \eqref{consmod} is a more standardised 
and generalised formulation of Hamiltonian mechanics.

As an additional note, one can modify the Lagrangian 
\eqref{rflag} to parametrisation wrt observed time 
$x^0 = t$, by writing $g_{00} (\bm x) = 1 - 2 \Phi (\bm x)$, 
setting $\dot t = 1 \ \Rightarrow \ d \tau = d t$. 
Upon binomial expansion of the part under square root 
up to the first order for non-relativistic approximation, 
we get as shown in \cite{cg}:
\begin{equation} \label{classlag} 
     \begin{split}
          L = 
          \frac12 
          g_{ij} (\bm x) 
          \dot x^i 
          \dot x^j + 
          \mathbb A_i (\bm x) 
          \dot x^i - 
          V (\bm x) 
     \end{split}, 
\qquad \text{where } \ 
     \begin{split} 
          \mathbb A_i (\bm x) &= 
          A_i (\bm x) + 
          g_{0i} (\bm x) 
     \\  
          V (\bm x) &= 
          \Phi (\bm x) - 
          A_0 (\bm x)
     \end{split}. 
\end{equation} 
If $A_\mu (\bm x) = 0$ in \eqref{rflag}, then one essentially 
reproduces the natural Lagrangian form \eqref{natlag} 
under the setting $\dot t = 1$, showing that the natural 
Lagrangian \eqref{natlag} is more relevant in non-relativistic 
settings.

\numberwithin{equation}{section}

\section{The Jacobi-Maupertuis-Randers-Finsler (JMRF) metric} 
\label{sec:jacobi}

The projection of a geodesic onto a constant energy 
hypersurface which is described by the Jacobi metric 
is achieved by dismissing the time co-ordinate as a degree 
of freedom while preserving Hamilton's equations 
of motion. 
To do this, one simply dismisses the energy which 
is the canonical momentum paired to time as a constant 
of motion. 
\smallskip 

First, I shall discuss how to project the generalised RF 
metric \eqref{rfmod} in $n+1$ spacetime into the JMRF 
metric in $n$ spatial co-ordinates \cite{cggmw} via a new 
constraint with a conformal factor derived from the original 
constraint. 
Consider the following RF spacetime metric \eqref{rfmod} 
rewritten as: 
\begin{equation} \label{randermet} 
     d s_{RF} = 
     \sqrt{
          \kappa 
          d y^2 +  
          \gamma_{ij} (\bm x) 
          d x^i 
          d x^j + 
          g_{00} (\bm x) 
          \left( 
               d t + 
               \frac{g_{i0} (\bm x)}{g_{00} (\bm x)} 
               d x^i 
          \right)^2
     } + 
     A_i (\bm x) 
     d x^i + 
     A_0 (\bm x) 
     d t, 
\end{equation} 
where the spatial metric $\gamma_{ij} (\bm x)$ is given by:
\begin{equation} \label{spatmetric} 
     \gamma_{ij} (\bm x) = 
     g_{ij} (\bm x) - 
     \frac{g_{i0} (\bm x) g_{j0} (\bm x)}{g_{00} (\bm x)}. 
\end{equation} 
and the inverse of the spatial metric $\gamma_{ij} (\bm x)$ 
according to \eqref{spatmetric} is given by:
\[ \begin{split} 
     \gamma_{ij} (\bm x) 
     g^{jk} (\bm x) &= 
     g_{ij} (\bm x) 
     g^{jk} (\bm x) - 
     \frac{g_{i0} (\bm x)}{g_{00} (\bm x)} 
     \left( 
          g_{0j} (\bm x) 
          g^{jk} (\bm x) 
     \right) 
\\ 
     &=
     g_{ij} (\bm x) 
     g^{jk} (\bm x) - 
     \frac{g_{i0} (\bm x)}{g_{00} (\bm x)} 
     \left( 
          - g_{00} (\bm x) 
          g^{0k} (\bm x) 
     \right) = 
     g_{ij} (\bm x) 
     g^{jk} (\bm x) + 
     g_{i0} (\bm x) 
     g^{0k} (\bm x) = 
     \delta^k_i
\end{split} \] 
\begin{equation} \label{blockdiag} 
     \gamma_{ij} (\bm x) 
     g^{jk} (\bm x) = 
     \delta^k_i
\qquad \Rightarrow \qquad 
     \left[ 
          \gamma_{ij} (\bm x) 
     \right]^{(-1)} = 
     g^{ij} (\bm x), 
\end{equation} 
The canonical momenta are deduced according 
to \eqref{eulag}, where the momentum canonically 
paired with the time $x^0 = t$ is a constant of motion 
$p_0 = q$. 
The gauge-covariant momenta $\pi_\mu$ given 
by \eqref{rfgcmom1} for 
$d \sigma = 
\sqrt{\kappa d y^2 + d \Sigma^2}$, 
and $d \Sigma^2 = 
g_{\alpha \beta} (\bm x) d x^\alpha d x^\beta$ 
are: 
\begin{equation} \label{rfgcmom2} 
     \begin{split} 
          \pi_0 &= 
          q - A_0 (\bm x) = 
          g_{i0} (\bm x) 
          \frac{d x^i}{d \sigma} + 
          g_{00} (\bm x) 
          \frac{d t}{d \sigma}  = 
          Q (\bm x), 
     \\ 
          \pi_i &= 
          p_i - 
          A_i (\bm x) = 
          \gamma_{ij} (\bm x) 
          \frac{d x^j}{d \sigma} + 
          \frac{g_{i0} (\bm x)}{g_{00} (\bm x)} 
          Q (\bm x). 
     \end{split}
\end{equation} 
Here I will introduce a new gauge-covariant momentum 
$\bm \Pi$ from \eqref{rfgcmom2}:
\begin{equation} \label{newgcmom} 
     \Pi_i = 
     p_i - 
     \alpha_i (\bm x) = 
     \gamma_{ij} (\bm x) 
     \frac{d x^j}{d \sigma} 
\qquad \Rightarrow \qquad 
     \frac{d x^i}{d \sigma} = 
     g^{ij} (\bm x) 
     \Pi_j, 
\qquad \text{where } \ 
     \alpha_i (\bm x) = 
     A_i (\bm x) + 
     \frac{g_{i0} (\bm x)}{g_{00} (\bm x)} 
     Q (\bm x)
\end{equation} 
and we must remember that since 
$\gamma_{ij} (\bm x) = 
g_{ij} (\bm x) - 
\dfrac{g_{i0} (\bm x) g_{j0} (\bm x)}{g_{00} (\bm x)}$
\[ \begin{split} 
     \gamma_{ij} (\bm x) 
     g^{jk} (\bm x) &= 
     g_{ij} (\bm x) 
     g^{jk} (\bm x) - 
     \frac{g_{i0} (\bm x)}{g_{00} (\bm x)} 
     \left( 
          g_{0j} (\bm x) 
          g^{jk} (\bm x) 
     \right) 
\\ 
     &=
     g_{ij} (\bm x) 
     g^{jk} (\bm x) - 
     \frac{g_{i0} (\bm x)}{g_{00} (\bm x)} 
     \left( 
          - g_{00} (\bm x) 
          g^{0k} (\bm x) 
     \right) = 
     g_{ij} (\bm x) 
     g^{jk} (\bm x) + 
     g_{i0} (\bm x) 
     g^{0k} (\bm x) = 
     \delta^k_i
\end{split} \] 
\begin{equation} \label{blockdiag} 
     \gamma_{ij} (\bm x) 
     g^{jk} (\bm x) = 
     \delta^k_i
\qquad \Rightarrow \qquad 
     \left[ 
          \gamma_{ij} (\bm x) 
     \right]^{(-1)} = 
     g^{ij} (\bm x). 
\end{equation} 
Thus, using the gauge covariant momentum $\Pi$ 
form \eqref{newgcmom} the constraint \eqref{consmod} 
for the RF metric\eqref{randermet} is written as: 
\begin{equation} \label{constr} 
     \phi (\bm x, \bm p) = 
     \sqrt{
          p_y^2 +  
          g^{ij} (\bm x) 
          \Pi_i \Pi_j + 
          \frac{( Q (\bm x) )^2}{g_{00} (\bm x)}
     } = 1, 
\qquad \text{where } \ 
     p_y = 
     \begin{cases} 
          0 \ , \ d \Sigma^2 \neq 0
     \\ 
          1 \ , \ d \Sigma^2 = 0
     \end{cases}.
\end{equation} 
To formulate the Jacobi metric, I shall rewrite 
the constraint \eqref{constr} for the RF metric 
\eqref{randermet} with time into 
a different constraint of the same form as 
\eqref{constraint} 
$$
     g^{ij} (\bm x) 
     \Pi_i 
     \Pi_j = 
     1 - 
     p_y^2 - 
     \frac{( Q (\bm x) )^2}{g_{00} (\bm x)}, 
$$
\begin{equation} \label{jcons} 
\Rightarrow \qquad 
     \Gamma (\bm x, \bm p) = 
     \sqrt{
          \left( 
               1 - 
               p_y^2 - 
               \frac{( Q (\bm x) )^2}{g_{00} (\bm x)} 
          \right)^{-1} 
          g^{ij} (\bm x) \Pi_i \Pi_j
     } = 1, 
\qquad \text{where } \ 
     p_y = 
     \begin{cases} 
          0 \ , \ d \Sigma^2 \neq 0
     \\ 
          1 \ , \ d \Sigma^2 = 0
     \end{cases}. 
\end{equation} 
We can therefore define the Jacobi metric as:
\begin{equation} \label{jmet} 
     J^{ij} (\bm x) := 
     \left( 
          1 - 
          p_y^2 -  
          \frac{( Q (\bm x) )^2}{g_{00} (\bm x)} 
     \right)^{-1} 
     g^{ij} (\bm x)  
\qquad \Rightarrow \qquad 
     J_{ij} (\bm x) = 
     \left( 
          1 - 
          p_y^2 - 
          \frac{( Q (\bm x) )^2}{g_{00} (\bm x)} 
     \right) 
     \gamma_{ij} (\bm x).
\end{equation} 
Upon applying \eqref{jmet} to the constraint \eqref{jcons}, 
the 1st of the constraint equations \eqref{conseq} allows 
us to write:
\begin{equation} \label{jmom} 
     \frac{d x^i}{d \lambda} = 
     \frac{\partial \Gamma}{\partial p_i} = 
     J^{ij} (\bm x) 
     \Pi_j 
\qquad \Rightarrow \qquad 
     \Pi_i = 
     p_i - 
     \alpha_i (\bm x) = 
     J_{ij} (\bm x) 
     \frac{d x^j}{d \lambda},
\end{equation} 
\begin{equation} \label{jpar} 
     J^{ij} (\bm x) 
     \Pi_i 
     \Pi_j = 
     J_{ij} (\bm x) 
     \frac{d x^i}{d \lambda} 
     \frac{d x^j}{d \lambda} = 1 
\qquad \Rightarrow \qquad 
     d \lambda^2 = 
     J_{ij} (\bm x) 
     d x^i 
     d x^j. 
\end{equation}
According to the Maupertuis principle \eqref{maup}, 
using \eqref{newgcmom}, \eqref{jmom}, and \eqref{jpar}, 
the JMRF metric can finally be written as: 
$$ 
     p_i = 
     J_{ij} (\bm x) 
     \frac{d x^j}{d \lambda} + 
     \alpha_i (\bm x)
$$ 
$$
\Rightarrow \qquad 
     d s_J = 
     p_i \; 
     d x^i = 
     \frac{J_{ij} (\bm x) d x^i d x^j}{d \lambda} + 
     \alpha_i (\bm x) 
     d x^i = 
     \sqrt{
          J_{ij} (\bm x) 
          d x^i 
          d x^j
     } + 
     \alpha_i (\bm x) 
     d x^i,
$$ 
\begin{equation} \label{rfjmet} 
     \boxed{
     d s_J = 
     \sqrt{
          \left( 
               1 - 
               p_y^2 -  
               \frac{( Q (\bm x) )^2}{g_{00} (\bm x)} 
          \right) 
          \gamma_{ij} (\bm x) 
          d x^i 
          d x^j
     } + 
     \left( 
          A_i (\bm x) + 
          \frac{g_{0i} (\bm x)}{g_{00} (\bm x)} 
          Q (\bm x) 
     \right) 
     d x^i, 
\quad \text{where } \ 
     p_y = 
     \begin{cases} 
          0 \ , \ \text{massive}
     \\ 
          1 \ , \ \text{light}
     \end{cases}.}
\end{equation} 
Furthermore, we can see from \eqref{newgcmom}, \eqref{jmet} 
and \eqref{jmom} that:
$$
     \frac{d x^i}{d \lambda} = 
     J^{ij} (\bm x) 
     \Pi_j = 
     \left( 
          1 - 
          p_y^2 - 
          \frac{( Q (\bm x) )^2}{g_{00} (\bm x)} 
     \right)^{-1} 
     g^{ij} (\bm x) 
     \Pi_j = 
     \left( 
          1 - 
          p_y^2 -  
          \frac{( Q (\bm x) )^2}{g_{00} (\bm x)} 
     \right)^{-1} 
     \frac{d x^i}{d \sigma}, 
$$
showing that the direction ratios along the geodesic 
of the Jacobi metric will be the same as with the original 
geodesic. 
$$
     \frac{d x^i}{d x^j} = 
     \frac{\dfrac{d x^i}{d \lambda}}{\dfrac{d x^j}{d \lambda}} = 
     \frac{\dfrac{d x^i}{d \sigma}}{\dfrac{d x^j}{d \sigma}}. 
$$ 
The JMRF described by \eqref{rfjmet} is a more 
complete and updated version of the Jacobi metric 
discussed previously in \cite{cgg1,cgg2,ilgm,cggmw}. 
Where previously in \cite{cggmw} my collaborators and 
I discussed the JMRF only for massive particles travelling 
along time-like curves in Riemannian spacetime metrics, 
this version is generalised to derive from RF spacetime 
metrics and accounts for the possibility of dealing with 
photons travelling along light-like or null curves. 
Special settings for massive particles and the new case 
of light-like null curves will be further discussed in the 
subsections to follow.

If the RF Lagrangian is parametrised wrt the cyclic 
co-ordinate and expanded binomially as shown in 
Section \ref{sec:prelim}, but without restriction to first 
order for low energy approximation, the Jacobi metric 
may be formulated as described by Maraner in 
\cite{maraner} for general Lagrangian systems. 
If more than one cyclic co-ordinate is available, 
the procedure can be repeated, until none are left.

\numberwithin{equation}{subsection}

\subsection{JMRF for massive particle under different settings} 
\label{sec:settings}

For the case of a massive particle, the JMRF 
is given by setting $p_y = 0$ in \eqref{rfjmet} 
according to \eqref{extramom}: 
\begin{equation} \label{massjmet} 
     d s_J = 
     \sqrt{
          \left( 
               1 - 
               \frac{( Q (\bm x) )^2}{g_{00} (\bm x)} 
          \right) 
          \gamma_{ij} (\bm x) 
          d x^i 
          d x^j
     } + 
     \left( 
          A_i (\bm x) + 
          \frac{g_{0i} (\bm x)}{g_{00} (\bm x)} 
          Q (\bm x) 
     \right) 
     d x^i.
\end{equation} 
Having formulated the Jacobi metric for the RF metric, 
I shall now discuss the Jacobi metric for different settings 
of the original RF metric.
\bigskip 

\subsubsection*{Riemannian metric}

If we start with a stationary Riemannian metric 
by setting $A_\mu (\bm x) = 0$ in \eqref{rfmetric}, 
then we will have the gauge fields according 
to \eqref{rfgcmom2} and \eqref{newgcmom}:
$$
     Q (\bm x) = q 
, \qquad 
     \alpha_i (\bm x) = 
     - \gamma_{ik} (\bm x) g^{k0} (\bm x) q,
$$ 
$$
     d s^2 = 
     g_{ij} (\bm x) 
     d x^i 
     d x^j + 
     2 g_{i0} (\bm x) 
     d x^i 
     d t + 
     g_{00} (\bm x) 
     d t^2
$$ 
and thus, we will have the Jacobi metric 
according to \eqref{rfjmet}:
\begin{equation} \label{jmriem} 
     d s_J = 
     \sqrt{
          \left( 
               1 - \frac{q^2}{g_{00} (\bm x)} 
          \right) 
          \gamma_{ij} (\bm x) 
          d x^i d x^j
     } + 
     q 
     \frac{g_{0i} (\bm x)}{g_{00} (\bm x)} 
     d x^i,
\end{equation} 
which is the result presented in \cite{cggmw}. 
This setting shows how the Jacobi metric creates new 
potentials from the metric.

\subsubsection*{Static spacetime with flat time component in non-relativistic limit}

In this case where $g_{00} (\bm x) = 1 , 
g_{0i} (\bm x) = 0 , A_\mu (\bm x) \neq 0$, 
such that the spacetime metric can be written as: 
\begin{equation} \label{flatmetric} 
     d s = 
     \sqrt{
          d t^2 - 
          g_{ij} (\bm x) 
          d x^i 
          d x^j
     } + 
     A_i (\bm x) 
     d x^i + 
     A_0 (\bm x) 
     d t, 
\end{equation} 
if we consider the non-relativistic limit by approximating 
up to first order of the binomial series expansion of the 
square root term of \eqref{flatmetric}, we will have: 
\begin{equation} \label{nrlimit} 
     d s = 
     d t \; 
     L \left( \bm{x}, \frac{d \bm{x}}{d t} \right) \approx 
     d t 
     \left[ 
          1 - 
          \left( 
               \frac12 
               g_{ij} (\bm x) 
               \frac{d x^i}{d t} 
               \frac{d x^j}{d t} - 
               A_i (\bm x) 
               \frac{d x^i}{d t} - 
               A_0 (\bm x)
          \right) 
     \right] 
\end{equation} 
which is the form of the action with the classical Lagrangian 
\eqref{classlag}. 
Then the Jacobi metric of \eqref{flatmetric} according 
to \eqref{rfjmet} is:
\begin{equation} \label{regjmrf} 
     d s_J = 
     \sqrt{
          \left( 
               1 - 
               \left( Q (\bm x) \right)^2 
          \right) 
          g_{ij} (\bm x) 
          d x^i 
          d x^j
     } + 
     A_i (\bm x) \ 
     d x^i, 
\qquad \text{where } \ 
     Q (\bm x) = 
     q - A_0 (\bm x).
\end{equation} 
If we can say that 
the total relativistic energy $q$ is given 
by $q = 1 + \varepsilon$ where $\varepsilon$ 
is the total mechanical energy, then we can write 
for low energy limits 
$$
     \varepsilon - A_0 (\bm x) \ll 1 
\qquad \Rightarrow \qquad 
     \left( Q (\bm x) \right)^2 = 
     \left( 
          1 + 
          \varepsilon - A_0 (\bm x)
     \right)^2 
     \approx 
     1 + 
     2 \left( 
          \varepsilon - 
          A_0 (\bm x) 
     \right). 
$$ 
Thus, the flat Jacobi metric \eqref{regjmrf} 
is written as: 
\begin{equation} \label{nrjmrf} 
     \therefore \qquad 
     d s_J = 
     \sqrt{
          - 2 \left( \varepsilon - A_0 (\bm x) \right) 
          g_{ij} (\bm x) 
          d x^i d x^j
     } + 
     A_i (\bm x) \ d x^i,
\end{equation} 
which is the non-relativistic limit discussed 
in \cite{cgg1,cgg2} when $A_i (\bm x) = 0$ 
in \eqref{nrjmrf}, and in \cite{cggmw}, 
thus reproducing the familiar form of the Jacobi 
metric calculated for simple classical mechanical systems 
\cite{ong,shs,gibbons,ilgm}. 
Under the circumstances that 
$A_i (\bm x) = 0 \ \forall \ i$, 
this example describes a case where the Jacobi 
metric geometrises the potential $A_0 (\bm x)$ 
absorbing it into the metric.

\subsection{Jacobi metric for Null curves} 
\label{sec:null}

In 1662, Fermat speculated in his principle of least time 
\cite{bornwolf} that light travels along paths requiring the 
shortest time interval, defined by null-geodesics. 
This makes Fermat's principle the optical version of the 
Brachistochrone problem \cite{bernoulli1,bernoulli2}, 
as discussed by Erlichson \cite{erlichson} and Broer 
\cite{broer}. 
Null-geodesics are unique since the speed of a particle 
(photon) travelling along them remains unchanged under 
local Lorentz transformations. In special relativity, in flat 
spaces this leads to Einstein's postulate on the universality 
of the speed of light in all inertial frames, which holds true 
locally, even in refracting media. 

\noindent 
Since the length of a null curve vanishes, one may introduce 
a metric based on Fermat's principle, called the optical metric. 
We shall demonstrate this starting with the stationary spacetime 
metric given below: 
\begin{equation} \label{stationary} 
      ds^2 = 
      g_{00} (\bm x) dt^2 + 
      2 g_{i0} (\bm x) \ d t \ dx^i + 
      g_{ij} (\bm x) dx^i dx^j.
\end{equation} 
Viewing the null version of \eqref{stationary} as a quadratic equation 
$$
     g_{00} (\bm x) dt^2 + 
     2 g_{i0} (\bm x) \ d t \ dx^i + 
     g_{ij} (\bm x) dx^i dx^j = 0,
$$
we can write the optical metric $d s_\mathcal O$ 
as a solution of the quadratic equation for $dt$ 
$$
     dt = 
     \pm 
     \sqrt{
          - \frac{\gamma_{ij} (\bm x)}{g_{00} (\bm x)} 
          d x^i d x^j
     } - 
     \frac{g_{i0} (\bm x)}{g_{00} (\bm x)} dx^i, 
\qquad \text{where } \ 
     \gamma_{ij} (\bm x) = 
     g_{ij} (\bm x) - 
     \frac{g_{i0} (\bm x) g_{j0} (\bm x)}{g_{00} (\bm x)},
$$ 
where we will take $+$ solution since $dt > 0$
\begin{equation} \label{optical} 
     d s_{\mathcal O} = 
     dt = 
     \sqrt{
          - \frac{\gamma_{ij} (\bm x)}{g_{00} (\bm x)} 
          d x^i d x^j
     } - 
     \frac{g_{i0} (\bm x)}{g_{00} (\bm x)} dx^i,
\end{equation}
which we can see is a Randers type of Finsler metric 
\cite{Randers}. 
Maupertuis speculated in \cite{maup} that light passing 
through a medium was refracted due to gravitational effects, 
implying that from an optical perspective, one can interpret 
gravitational fields as transparent media and vice versa. 
\bigskip

\noindent 
Under the circumstances that one is dealing 
with a null curve, the JMRF is given by setting 
$p_y = 1$ in \eqref{rfjmet}: 
\begin{equation} \label{rfjmetmod} 
     d s_J = 
     \sqrt{ 
          - \frac{( Q (\bm x) )^2}{g_{00} (\bm x)} 
          \gamma_{ij} (\bm x) d x^i d x^j
     } + 
     \left( 
          A_i (\bm x) + 
          \frac{g_{0i} (\bm x)}{g_{00} (\bm x)} 
          Q (\bm x) 
     \right) d x^i.
\end{equation} 
Now, if we consider only Riemannian stationary 
metrics ($A_\mu (\bm x) = 0$ in \eqref{rfmetric}) 
then we will have the JMRF: 
\begin{equation} \label{riemjmetmod} 
     d \widetilde s_J = 
     \frac{d s_J}q = 
     \sqrt{ 
          - \frac{\gamma_{ij} (\bm x)}{g_{00} (\bm x)}  
          d x^i d x^j
     } + 
     \frac{g_{0i} (\bm x)}{g_{00} (\bm x)} d x^i.
\end{equation} 
Furthermore, if one is dealing with a static metric 
($g_{i0} (\bm x) = 0$) we can write: 
\begin{equation} \label{statjmetmod} 
     d \widetilde s_J = 
     \frac{d s_J}q = 
     \sqrt{ 
          - \frac{g_{ij} (\bm x)}{g_{00} (\bm x)}  
          d x^i d x^j}.
\end{equation} 
which happens to be the form of the optical metric 
formulated for static metric according to Fermat's 
principle. 
However, we can see from \eqref{riemjmetmod} that 
for stationary metrics the form of the Jacobi metric 
deviates from the optical metric \eqref{optical}. 
Thus, we can say that optical metrics are not Jacobi 
metrics for null curves, and their similarity for Riemannian 
static metrics as shown by \eqref{statjmetmod} is merely 
coincidence.

\numberwithin{equation}{section}

\section{Frame dragging effect} 
\label{sec:frame_drag} 

\noindent 
So far, we have seen that since there is no 
Hamiltonian described for the Jacobi metric, 
it is not always possible to describe mechanics 
using the Hamiltonian \eqref{ham} and 
Hamilton's equations \eqref{hameom}. 
Thus, the constraint \eqref{constraint} 
and its equations \eqref{conseq} are more 
reliable alternatives to Hamilton's equations. 
This is furthermore evident when one considers 
the question of frame dragging in the Jacobi metric. 
In \cite{epstein}, Epstein discusses a Hamiltonian 
approach to studying frame dragging. 
Such frame dragging effects should also exist in 
a Jacobi metric based description of dynamics 
around a black hole. 
However, since a Hamiltonian is unavailable, 
it should also be possible to describe frame 
dragging using the constraint. 
\smallskip 

Here, I will discuss the frame dragging effect 
in the context of the Jacobi metric for Riemannian 
and RF spacetime metrics separately.

\numberwithin{equation}{subsection} 

\subsection{Riemannian metric} 
\label{sec:riemann} 

According to Epstein \cite{epstein}, the frame 
dragging effect describes motion independent 
of momentum. 
However, if we use the constraint instead of the 
Hamiltonian, then the constraint for stationary 
metric \eqref{stationary} according to 
\eqref{constraint} is given by: 
\begin{align} 
          \phi (\bm{x, p}) &= 
          \sqrt{
               p_y^2 + 
               g^{ij} (\bm x) 
               p_i p_j + 
               2 g^{i0} (\bm x) 
               p_i p_0 + 
               g^{00} (\bm x) 
               (p_0)^2 
          } 
     \nonumber 
     \\ 
     \label{statcons} 
          &= \sqrt{
               p_y^2 + 
               f^{ij} (\bm x) 
               p_i p_j + 
               g^{00} (\bm x) 
               \left( 
                    p_0 + 
                    \frac{g^{0m} (\bm x)}{g^{00} (\bm x)} 
                    p_m 
               \right)^2 
          } = 1, 
\end{align} 
where 
$f^{ij} (\bm x) = 
g^{ij} (\bm x) - 
\frac{g^{i0} (\bm x) g^{j0} (\bm x)}{g^{00} (\bm x)}$ 
is the spatial inverse metric. 
We shall have the following constraint equations 
according to \eqref{conseq}: 
\begin{align} 
          \frac{d x^i}{d s} &= 
          \frac{\partial \phi}{\partial p_i} = 
          g^{ij} (\bm x) p_j + 
          g^{i0} (\bm x) p_0 = 
          f^{ij} (\bm x) p_j + 
          \frac{g^{i0} (\bm x)}{g^{00} (\bm x)} 
          \left( 
               g^{00} (\bm x) 
               p_0 + 
               g^{0j} (\bm x) 
               p_j
          \right), 
     \nonumber 
     \\ 
     \label{statconseq} 
          \frac{d t}{d s} &= 
          \frac{\partial \phi}{\partial p_0} = 
          g^{0j} (\bm x) p_j + 
          g^{00} (\bm x) p_0 = 
          \sqrt{
               g^{00} (\bm x) 
               \left( 
                    1 - 
                    p_y^2 - 
                    f^{ij} (\bm x) 
                    p_i 
                    p_j 
               \right) 
          }
\end{align} 
from which we can see that 
\begin{equation} \label{vel} 
     \frac{d x^i}{d t} = 
     \frac{f^{ij} (\bm x)}{\sqrt{g^{00} (\bm x) \left( 1 - p_y^2 - f^{ij} (\bm x) p_i p_j \right)}} 
     p_j + 
     \frac{g^{i0} (\bm x)}{g^{00} (\bm x)}, 
\end{equation} 
which matches what Epstein discussed in \cite{epstein}. 
Here we must note that if we consider massless 
particles for null curves by setting $p_y = 1$, then 
the drift velocity under frame-dragging effect cannot 
be determined due to an emerging singularity upon 
setting $p_i = 0$ in \eqref{vel}. 
This is consistent with the fact that light cannot 
be described with zero spatial momentum because 
if we set $p_i = 0$ and $p_y = 1$ in the constraint 
\eqref{statcons}, then we must also have $p_0 = 0$. 
\begin{equation} \label{limit} 
     p_i = 0, \ p _y = 1 
\qquad \Rightarrow \qquad      
     \phi (\bm{x, p}) = 
     \sqrt{1 + g^{00} (\bm x) p_0^2} = 1 
\qquad \Rightarrow \qquad      
     p_0 = 0.
\end{equation} 
Thus, from here on, I shall discuss frame-dragging 
effect for massive particles only where $p_y = 0$. 
Ultimately, frame dragging is manifested as motion 
in the form of a drift velocity $D^i (\bm x)$ that exists 
in the absence of spatial momentum. 
This can also be seen from \eqref{vel} or directly from 
the constraint equations \eqref{statconseq} 
for massive particles $p_y = 0$: 
\begin{equation} \label{fdrag} 
     p_j = 0 \quad \forall \ j 
\ \Rightarrow \ 
     \left\{ \begin{split} 
          \left( 
               \frac{d x^i}{d s} 
          \right)_{p_j = 0} &= 
          g^{i0} (\bm x) 
          p_0 
     \\ 
          \left( 
               \frac{d t}{d s} 
          \right)_{p_j = 0} &= 
          g^{00} (\bm x) 
          p_0    
     \end{split} \right\} 
\ \Rightarrow \ 
     D^i (\bm x) = 
     \left( 
          \frac{d x^i}{d t} 
     \right)_{p_j = 0} = 
     \frac{g^{i0} (\bm x)}{g^{00} (\bm x)}, 
\end{equation} 
where, we can see that the background drift 
velocity $D^i (\bm x)$ manifesting due to the frame 
dragging effect is completely a function of the inverse 
spacetime metric. 

Considering the JMRF for stationary 
Riemannian spacetime metrics described by setting 
$A_i (\bm x) = 0$ in \eqref{rfjmet} and remembering 
\eqref{fdrag} and that 
$g_{i0} (\bm x) g^{00} (\bm x) = - g_{ij} (\bm x) g^{j0} (\bm x)$, 
we can encode the drift velocity $D^i (\bm x)$ from 
the frame dragging effect into the Jacobi metric using 
\eqref{substitute} as shown below: 
\begin{equation} \label{encoderiem} 
     d s_{J (R)} = 
     \sqrt{
          \left( 
               1 - 
               p_y^2 -  
               \frac{q^2}{g_{00} (\bm x)} 
          \right) 
          \gamma_{ij} (\bm x) 
          d x^i 
          d x^j
     } - 
     q 
     D_i (\bm x) 
     d x^i, 
\quad \text{where } \ 
     p_y = 
     \begin{cases} 
          0 \ , \ \text{massive}
     \\ 
          1 \ , \ \text{light}
     \end{cases}
\end{equation} 
where we have 
$D_i (\bm x) = g^{00} (\bm x) \gamma_{ij} (\bm x) D^j (\bm x)$. 
From the constraint \eqref{statcons}, 
we can say that: 
\begin{equation} \label{nullmom} 
     p_i = 0, p_y = 0 
\qquad \Rightarrow \qquad 
     g^{00} (\bm x) 
     \left( p_0 \right)^2 = 
     1,
\end{equation}  
which allows us to write the proper velocities 
of \eqref{fdrag} as: 
\begin{equation} \label{fdragvel} 
     \left( 
          \frac{d x^i}{d s} 
     \right)_{p_j = 0} = 
     \frac{g^{i0} (\bm x)}{\sqrt{g^{00} (\bm x)}} 
\qquad , \qquad 
     \left( 
          \frac{d t}{d s} 
     \right)_{p_j = 0} = 
     \sqrt{g^{00} (\bm x)}. 
\end{equation} 
Most importantly, we have the frame dragging 
proper velocity given by the first equation 
of \eqref{fdragvel}, which is completely independent 
of momentum.  
Consider the Jacobi metric for a stationary Riemannian 
metric given by \eqref{jmriem}. 
This metric has the constraint given according to \eqref{jcons}: 
\begin{equation} \label{jmriemcons}
     \Gamma (\bm x, \bm p) = 
     \sqrt{
          \left( 
               1 - 
               \frac{q^2}{\left( V (\bm x) \right)^2} 
          \right)^{-1} 
          g^{ij} (\bm x) 
          \left( 
                p_i - 
                q 
                W_i (\bm x) 
          \right) 
          \left( 
                p_j - 
                q 
                W_j (\bm x) 
          \right) 
     } = 1,  
\end{equation} 
where we have: 
\begin{equation} \label{substitute} 
     W_i (\bm x) = 
     \frac{g_{0i} (\bm x)}{\left( V (\bm x) \right)^2} 
\qquad , \qquad 
     \left( V (\bm x) \right)^2 = 
     g_{00} (\bm x), 
\end{equation} 
from which according to the constraint equations for Jacobi 
metric \eqref{jmom}, we will have: 
\begin{equation} \label{jmriemeom} 
     \frac{d x^i}{d \lambda} = 
     \frac{\partial \Gamma}{\partial p_i} = 
     \left( 
          1 - 
          \frac{q^2}{\left( V (\bm x) \right)^2} 
     \right)^{-1} 
     g^{ij} (\bm x) 
     \left( 
           p_j - 
           q 
           W_j (\bm x) 
     \right). 
\end{equation} 
Thus, as with \eqref{fdrag}, we can describe the frame 
dragging velocity to be: 
\begin{equation} \label{jmriemfdrag1} 
     \left( 
          \frac{d x^i}{d \lambda} 
     \right)_{p_j = 0} = 
     \left( 
          1 - 
          \frac{q^2}{\left( V (\bm x) \right)^2} 
     \right)^{-1} 
     q 
     W^i (\bm x), 
\qquad \text{where } \      
     W^i (\bm x) = 
     - g^{ij} (\bm x) 
     W_j (\bm x) = 
     g^{i0} (\bm x).
\end{equation} 
Furthermore, from the constraint \eqref{jmriemcons}, 
we will have: 
$$
     \left( \Gamma (\bm x, \bm p) \right)_{p_i = 0} = 
     \sqrt{
          \left( 
               1 - 
               \frac{q^2}{\left( V (\bm x) \right)^2}
          \right)^{-1} 
          q^2 
          \left| W (\bm x) \right|^2   
     } = 
     1 
$$ 
\begin{equation} \label{conscond} 
\Rightarrow \qquad 
     q^2 
     \left( 
          \left| W (\bm x) \right|^2 + 
          \frac1{\left( V (\bm x) \right)^2}
     \right) = 
     1, 
\end{equation} 
where 
$\left| W (\bm x) \right|^2 = 
g^{ij} (\bm x) W_i (\bm x) W_j (\bm x) = 
\gamma_{ij} (\bm x) W^i (\bm x) W^j (\bm x) = 
- g^{0i} (\bm x) 
\frac{g_{i0} (\bm x)}{g_{00} (\bm x)}$, 
which allows us to write \eqref{jmriemfdrag1} as: 
\begin{equation} \label{jmriemfdrag2} 
     D^i_J (\bm x) = 
     \left( 
          \frac{d x^i}{d \lambda} 
     \right)_{p_j = 0} = 
     \frac1{q} 
     \frac{W^i (\bm x)}{\left| W (\bm x) \right|^2} = 
     \sqrt{
          \left| W (\bm x) \right|^2 + 
          \frac1{\left( V (\bm x) \right)^2} 
     } 
     \frac{W^i (\bm x)}{\left| W (\bm x) \right|^2}. 
\end{equation} 
Upon substituting the functions with \eqref{substitute}, 
we have from \eqref{jmriemfdrag2}: 
$$ 
     D^i_J (\bm x) = 
     \left( 
          1 - \frac{1}{g^{00} (\bm x) g_{00} (\bm x)} 
     \right)^{-1}  
     \frac{g^{i0} (\bm x)}{\sqrt{g^{00} (\bm x)}} = 
     \left( 
          1 - \frac{1}{g^{00} (\bm x) g_{00} (\bm x)} 
     \right)^{-1}  
     \sqrt{g^{00} (\bm x)} 
     D^i (\bm x). 
$$ 
\begin{equation} \label{jmriemfdrag3} 
     D^i (\bm x) = 
     \left( 
          \frac{V (\bm x)}{\sqrt{1 + \left| W (\bm x) \right|^2 \left( V (\bm x) \right)^2}} 
     \right)^3 
     \left| W (\bm x) \right|^2 
     D^i_J (\bm x). 
\end{equation} 
Thus, when the Hamiltonian is absent for cases 
such as the Jacobi metric, the constraint proves 
much more suitable for dynamical analysis. 
We can see that the background drift 
parametrised wrt the Jacobi metric $D^i_J (\bm x)$ 
can be described in terms of the JMRF data 
or the original metric, and the original drift velocity 
$D^i (\bm x)$ under frame dragging effect 
can be deduced from it as well.

\subsection{RF metric} 
\label{sec:randers-finsler} 

If we consider a general RF spacetime metric 
\eqref{rfmetric}, its constraint is given by simply 
replacing all the momenta $p_\mu$ with gauge-covariant 
momenta $\pi_\mu$ in \eqref{statcons}: 
\begin{equation} \label{rfcons} 
     \phi_{RF} (\bm{x, p}) = 
     \sqrt{
          p_y^2 + 
          f^{ij} (\bm x) 
          \pi_i \pi_j + 
          g^{00} (\bm x) 
          \left( 
               Q (\bm x) + 
               \frac{g^{0m} (\bm x)}{g^{00} (\bm x)} 
               \pi_m 
          \right)^2 
     } = 1, 
\end{equation} 
which leads to the following constraint equations 
similar to \eqref{statconseq}: 
\begin{align} 
          \frac{d x^i}{d \sigma} &= 
          \frac{\partial \phi_{RF}}{\partial p_i} = 
          f^{ij} (\bm x) \pi_j + 
          \frac{g^{i0} (\bm x)}{g^{00} (\bm x)} 
          \left( 
               g^{00} (\bm x) 
               Q (\bm x) + 
               g^{0j} (\bm x) 
               \pi_j
          \right), 
     \nonumber 
     \\ 
     \label{rfconseq} 
          \frac{d t}{d \sigma} &= 
          \frac{\partial \phi_{RF}}{\partial p_0} = 
          \sqrt{
               g^{00} (\bm x) 
               \left( 
                    1 - 
                    p_y^2 - 
                    f^{ij} (\bm x) 
                    \pi_i 
                    \pi_j 
               \right) 
          }
\end{align} 
from which we can see that the velocity is given 
by the same replacement in \eqref{vel}. 
\begin{equation} \label{rfvel} 
     \frac{d x^i}{d t} = 
     \frac{f^{ij} (\bm x)}{\sqrt{g^{00} (\bm x) \left( 1 - p_y^2 - f^{ij} (\bm x) \pi_i \pi_j \right)}} 
     \pi_j + 
     \frac{g^{i0} (\bm x)}{g^{00} (\bm x)}. 
\end{equation} 
This time, the drift velocity $D^i_{RF} (\bm x)$ is given as: 
\begin{equation} \label{fdragrf} 
     D^i_{RF} (\bm x) = 
     \left( 
          \frac{d x^i}{d t} 
     \right)_{p_j = 0} = 
     - \frac{f^{ij} (\bm x) A_j (\bm x)}
     {\sqrt{g^{00} (\bm x) \left( 1 - p_y^2 - f^{ij} (\bm x) A_i (\bm x) A_j (\bm x) \right)}} + 
     \frac{g^{i0} (\bm x)}{g^{00} (\bm x)}, 
\end{equation} 
Thus, if we write the JMRF for general RF 
spacetime metrics \eqref{rfjmet}, we can 
encode the drift velocity \eqref{fdragrf} as: 
\begin{equation} \label{encoderf} 
     d s_{J (RF)} = 
     \sqrt{
          \left( 
               1 - 
               p_y^2 -  
               \frac{\left(Q (\bm x) \right)^2}{g_{00} (\bm x)} 
          \right) 
          \gamma_{ij} (\bm x) 
          d x^i 
          d x^j
     } + 
     \left( 
     A_i (\bm x) + 
     Q (\bm x) 
     D^{RF}_i (\bm x) 
     \right) 
     d x^i, 
\quad \text{where } \ 
     p_y = 
     \begin{cases} 
          0 \ , \ \text{massive}
     \\ 
          1 \ , \ \text{light}
     \end{cases}
\end{equation} 
for which $D^{RF}_i (\bm x)$ is given by: 
$$
     D^{RF}_i (\bm x) = 
     g^{00} (\bm x) 
     \gamma_{ij} (\bm x) 
     \left( 
          D^j_{RF} + 
          \frac{f^{jm} (\bm x) A_m (\bm x)}
          {\sqrt{g^{00} (\bm x) \left( 1 - p_y^2 - f^{ab} (\bm x) A_a (\bm x) A_b (\bm x) \right)}}
     \right).
$$
One difference as a result of exploring RF 
spacetime metrics as opposed to Riemannian 
metrics is that we are not restricted from 
describing the drift velocity for light-like 
null curves by setting $p_y = 1$ in \eqref{fdragrf} 
that might lead to a singularity. 
This is because in RF spacetime metrics, 
setting spatial momenta to vanish does not 
lead to vanishing energy, unlike what was 
seen with Riemannian spacetimes with 
\eqref{limit}. 
This drift velocity for a photons upon setting 
spatial momenta to vanish is thus given by 
setting $p_y = 1$ in \eqref{fdragrf}: 
\begin{equation} \label{fdragphoton} 
     \left( 
          D^i_{RF} (\bm x) 
     \right)_{p_y = 1} = 
     \frac{f^{ij} (\bm x) A_j (\bm x)}
     {\sqrt{g^{00} (\bm x) \left( f^{ij} (\bm x) A_i (\bm x) A_j (\bm x) \right)}} + 
     \frac{g^{i0} (\bm x)}{g^{00} (\bm x)}, 
\end{equation} 
Returning to the JMRF metric for general 
RF metrics of time-like curves of massive particles 
given by \eqref{rfjmet}, the constraint given according 
to \eqref{jcons}: 
\begin{equation} \label{jmriemcons2}
     \Gamma (\bm x, \bm p) = 
     \sqrt{
          \left( 
               1 - 
               \left( 
                    \frac{Q (\bm x)}{V (\bm x)} 
               \right)^2 
          \right)^{-1} 
          g^{ij} (\bm x) 
          \left( 
                p_i - 
                \Omega_i (\bm x) 
          \right) 
          \left( 
                p_j - 
                \Omega_j (\bm x) 
          \right) 
     } = 1,  
\end{equation} 
where we have: 
\begin{equation} \label{substitute2} 
     \Omega_i (\bm x) = 
     A_i (\bm x) + 
     Q (\bm x) 
     \frac{g_{0i} (\bm x)}{g_{00} (\bm x)}, 
\end{equation} 
from which according to the constraint equations for Jacobi 
metric \eqref{jmom}, we will have: 
\begin{equation} \label{jmriemeom2} 
     \frac{d x^i}{d \lambda} = 
     \frac{\partial \Gamma}{\partial p_i} = 
     \left( 
          1 - 
          \left( 
               \frac{Q (\bm x)}{V (\bm x)} 
          \right)^2 
     \right)^{-1} 
     g^{ij} (\bm x) 
     \left( 
           p_j - 
           \Omega_j (\bm x) 
     \right). 
\end{equation} 
Thus, as with \eqref{fdrag}, we can describe the frame 
dragging velocity to be: 
\begin{equation} \label{jmriemfdrag4} 
     \left( 
          \frac{d x^i}{d \lambda} 
     \right)_{p_j = 0} = 
     \left( 
          1 - 
          \left( 
               \frac{Q (\bm x)}{V (\bm x)} 
          \right)^2 
     \right)^{-1} 
     \left( 
          Q (\bm x) 
          W^i (\bm x) + 
          g^{ij} (\bm x) 
          A_j (\bm x)
     \right), 
\end{equation} 
which concludes our study of frame dragging effect 
for JMRF metrics derived for general RF spacetimes.

\numberwithin{equation}{section}

\section{Geometrising the RF metric} 
\label{sec:rfgeom}

The RF metric describes a relativistic system with potentials 
comparable to a magnetic gauge field, which leads one to ask 
if these potentials can be geometrised via Eisenhart lift 
in the same manner as in the usual non-relativistic systems 
it is applied to. 
\bigskip 

\noindent 
In the interest of briefly revisiting the procedure of the Eisenhart 
lift discussed in \cite{CarAlv}, suppose we have the natural 
Lagrangian \eqref{natlag} where $x^0 = t$ is a cyclic co-ordinate, 
written as: 
\begin{equation} \label{lag.eisenhart1} 
     L = 
     \frac12 
     g_{\mu \nu} (\bm x) 
     \dot x^\mu 
     \dot x^\nu = 
     \frac12 
     \gamma_{ij} (\bm x) 
     \dot x^i 
     \dot x^j + 
     \frac12 
     g_{00} (\bm x) 
     \left( 
          \dot t + 
          \frac{g_{i0} (\bm x)}{g_{00} (\bm x)} 
          \dot x^i
     \right)^2, 
\qquad \text{where } \      
     \gamma_{ij} (\bm x) = 
     g_{ij} (\bm x) - 
     \frac{g_{i0} (\bm x) g_{j0} (\bm x)}{g_{00} (\bm x)}. 
\end{equation} 
The regular Hamiltonian according to Legendre's principle 
\eqref{regham} is: 
\[ 
     \begin{split} 
          p_i &= 
          \frac{\partial L}{\partial \dot x^i} = 
          \gamma_{ij} (\bm x) 
          \dot x^j + 
          g_{i0} (\bm x) 
          \left( 
               \dot t + 
               \frac{g_{i0} (\bm x)}{g_{00} (\bm x)} 
               \dot x^i
          \right), 
     \\     
          p_0 &= 
          \frac{\partial L}{\partial \dot t} = 
          g_{00} (\bm x) 
          \left( 
               \dot t + 
               \frac{g_{i0} (\bm x)}{g_{00} (\bm x)} 
               \dot x^i
          \right)    
     \end{split} 
\]
\begin{equation} \label{ham.eisenhart1} 
     \mathcal H = 
     p_\mu 
     \dot x^\mu - 
     L = 
     \frac12 
     g^{ij} (\bm x) 
     \pi_i 
     \pi_j + 
     \frac{p^2_0}{2 g_{00} (\bm x)}, 
\qquad \text{where } \      
     \pi_i = 
     p_i - 
     \frac{g_{i0} (\bm x)}{g_{00} (\bm x)} 
     p_0. 
\end{equation} 
Upon writing $p_0 = q$, and 
$\mathbb A (\bm x) = \dfrac{g_{i0} (\bm x)}{g_{00} (\bm x)} q$, 
$V (\bm x) = \dfrac{q^2}{2 g_{00} (\bm x)}$, we will have 
\eqref{ham.eisenhart1} become: 
\begin{equation} \label{ham.eisenhart2} 
     p_0 = q 
\qquad \Rightarrow \qquad 
     \mathcal H = 
     \frac12 
     g^{ij} (\bm x) 
     \pi_i 
     \pi_j + 
     V (\bm x), 
\qquad \text{where } \      
     \pi_i = 
     p_i - 
     \mathbb A_i (\bm x) 
     q. 
\end{equation} 
According to Hamilton's equation of motion, 
from \eqref{ham.eisenhart2} 
\begin{equation} \label{eisenhart.eom} 
     \dot x^i = 
     \frac{\partial \mathcal H}{\partial p_i} = 
     g^{ij} (\bm x) 
     \pi_j 
\end{equation} 
we can use the Legendre's principle of \eqref{ham} 
in reverse to write the Lagrangian $\mathcal L$ 
by applying \eqref{eisenhart.eom}: 
$$
     \mathcal L = 
     p_i 
     \dot x^i - \mathcal H = 
     \frac12 
     g^{ij} (\bm x) 
     \pi_i 
     \pi_j + 
     g^{ij} (\bm x) 
     \pi_j 
     \mathbb A_i (\bm x) - 
     V (\bm x), 
$$
\begin{equation} \label{lag.eisenhart2} 
     \mathcal L = 
     \frac12 
     \gamma_{ij} (\bm x) 
     \dot x^i 
     \dot x^j + 
     \mathbb A_i (\bm x) 
     \dot x^i - 
     V (\bm x), 
\end{equation} 
which is the familiar classical Lagrangian \eqref{classlag}. 
Thus, we can say that the Lagrangian \eqref{lag.eisenhart1} 
is the Eisenhart lift of the Lagrangian \eqref{lag.eisenhart2} 
where the potentials $\mathbb A_i (\bm x)$ and $V (\bm x)$ 
have been geometrised by inserting the canonical pair $(t, q)$, 
which can be seen by applying the replacements 
$\mathbb A (\bm x) = 
\dfrac{g_{i0} (\bm x)}{g_{00} (\bm x)} q$, 
$V (\bm x) = 
\dfrac{q^2}{2 g_{00} (\bm x)}$ 
to \eqref{lag.eisenhart1}: 
\begin{equation} \label{lag.eisenhart3} 
     L = 
     \frac12 
     \left( 
          g_{ij} (\bm x) - 
          \frac{\mathbb A_i (\bm x) \mathbb A_j (\bm x)}{4 V (\bm x)} 
     \right) 
     \dot x^i 
     \dot x^j + 
     \frac{q}{2 V (\bm x)} 
     \mathbb A_i (\bm x) 
     \dot x^i 
     \dot t + 
     \frac{q^2}{4 V (\bm x)} 
     \dot t^2. 
\end{equation} 
However, when attempted directly, the Eisenhart lift faces 
limitations when dealing with RF metrics, so we must seek 
an alternative to extend its utility beyond natural Hamiltonian 
systems.

\numberwithin{equation}{subsection}

\subsection{Eisenhart lift via Jacobi metric} 
\label{sec:eisenhart}

So far, the Eisenhart lift was performed for natural Hamiltonian 
systems. Here, I shall attempt to do the same by using the constraint 
to project a curve described by a Riemannian metric onto a fixed 
hypersurface as a RF metric. 
Consider the Riemannian metric given below with cyclic co-ordinate 
$x^0 = T$: 
\begin{equation} \label{riemlag} 
     d s^2_R = 
     G_{ij} (\bm x) 
     d x^i 
     d x^j + 
     2 G_{i0} (\bm x) 
     d x^i 
     d T + 
     G_{00} (\bm x) 
     d T^2. 
\end{equation} 
Writing the Lagrangian according to \eqref{action} 
the canonical momenta from \eqref{riemlag} for 
$d \lambda = \sqrt{G_{\mu \nu} (\bm x) d x^\mu d x^\nu}$ 
and the conserved energy $k = const.$ are: 
$$
     L = 
     \sqrt{
          G_{ij} (\bm x) 
          \dot x^i 
          \dot x^j + 
          2 G_{i0} (\bm x) 
          \dot x^i 
          \dot T + 
          G_{00} (\bm x) 
          \dot T^2
     }
$$ 
\[ \begin{split} 
     p_T &= 
     G_{i0} (\bm x) 
     \frac{d x^i}{d \lambda} + 
     G_{00} (\bm x) 
     \frac{d T}{d \lambda} 
\\ 
     p_i &= \mathbb Y_{ij} (\bm x) \frac{d x^j}{d \lambda} + 
     k \frac{G_{i0} (\bm x)}{G_{00} (\bm x)}
\end{split}, 
\qquad  \text{where } \ 
     \mathbb Y_{ij} (\bm x) = 
     G_{ij} (\bm x) - \frac{G_{i0} (\bm x) G_{j0} (\bm x)}{G_{00} (\bm x)},
\]
using which I can define a gauge covariant momentum $\bm{\Pi^*}$: 
$$
     \Pi^*_i = 
     p_i - 
     k 
     \frac{G_{i0} (\bm x)}{G_{00} (\bm x)},
$$ 
and use it to write the constraint according to \eqref{constraint} as:
\begin{equation} \label{riemconstr} 
     \psi (\bm{x, p}) = 
     \sqrt{
          G^{ij} (\bm x) \Pi^*_i \Pi^*_j + 
          \frac{k^2}{G_{00} (\bm x)}
     } = 1 
\qquad , \qquad 
     \text{where } \ 
     G^{ik} (\bm x) 
     \mathbb Y_{kj} (\bm x) = 
     \delta^i_j.
\end{equation}
If we want the constraint \eqref{riemconstr} to match the form 
of \eqref{constraint}, the last term of \eqref{riemconstr} must vanish. 
In simple words, we require that
\begin{equation} \label{require} 
     \frac{k^2}{G_{00} (\bm x)} = 0.
\end{equation}
However, we cannot have $k = p_T = 0$, and the value 
of $G_{00} (\bm x)$ cannot be determined for \eqref{require} 
to hold. Thus, there is no way to rewrite the constraint 
$\psi (\bm{x, p})$ \eqref{riemconstr} for a Riemannian metric 
into the form for a lower dimensional constraint \eqref{constraint} 
for the RF metric. 
Conversely, it is not possible to directly Eisenhart lift a RF metric 
into a Riemannian metric by lifting the constraint.

Thus, while the Eisenhart lift is a proven method 
to insert additional directions of symmetry, it cannot 
be applied to RF metrics as shown above. 
While we are unable to directly lift a RF metric, 
there are some alternatives for geometrising the 
potentials of the linear term, sometimes under some 
conditions. 
One way is to convert the RF metric into a natural 
Lagrangian \eqref{natlag} by taking non-relativistic 
approximation and setting $\dot t = 1$. 
However, the linear terms then become merged with 
some of the metric components as seen in \eqref{classlag}, 
and the process is not reversible, unlike the Eisenhart lift 
from \eqref{lag.eisenhart2} to \eqref{lag.eisenhart3}.

On the other hand, the Jacobi metric is the reverse, ie.- 
converting geometry into potentials to hide a direction 
of symmetry applicable to RF geometry, so the reverse 
should prove a suitable alternative to Eisenhart lift. 
I will call this procedure the Eisenhart-Randers (ER) lift, 
which is applicable so long as one can identify a suitable 
conformal factor. 
\bigskip 

\noindent
Given a RF metric if we can identify 
a conformal factor in the metric such that 
\begin{equation} \label{org} 
     d s = 
     \sqrt{
          \left( 
               1 - \frac{(k - U (\bm x))^2}{\beta (\bm x)} 
          \right) 
          G_{ij} (\bm x) 
          d x^i d x^j
     } + 
     \left( 
          A_i (\bm x) + 
          \frac{\alpha_i (\bm x)}{\beta (\bm x)} 
          (k - U (\bm x)) 
     \right) 
     d x^i, 
\end{equation}
or in the constraint \eqref{jcons} such that 
\begin{equation} \label{confactor} 
     \Gamma (\bm{x, p}) = 
     \sqrt{
          \left( 
               1 - \frac{(k - U (\bm x))^2}{\beta (\bm x)} 
          \right)^{-1} 
          G^{ij} (\bm x) 
          \pi_i \pi_j
     }, 
\end{equation} 
$$
\text{where } \qquad 
     \pi_i = 
     p_i - 
     \left[ 
          A_i (\bm x) + 
          \frac{\alpha_i (\bm x)}{\beta (\bm x)} 
          \left( 
               k - U (\bm x) 
          \right) 
     \right],
$$
then by reversing the steps to derive JMRF 
metric \eqref{jmet}, I can deduce the ER metric 
by writing the constraint $\phi (\bm x, \bm p)$, 
lifting it (replacing $k = p_v$) and writing the first 
of constraint equations 
\eqref{conseq}.  
\begin{equation} \label{jlift} 
     \phi (\bm {x, p}) = 
     \sqrt{G^{ij} (\bm x) 
     \pi_i \pi_j + 
     \frac{\left(p_v - U (\bm x) \right)^2}{\beta (\bm x)}} = 
     \sqrt{
          \Omega_{\mu \nu} (\bm x) 
          \frac{d x^\mu}{d \theta} 
          \frac{d x^\nu}{d \theta}
     } = 1, 
\end{equation} 
\begin{align} 
          \frac{d v}{d \theta} &= 
          \frac{\partial \phi}{\partial p_v} = 
          \frac{p_v - U (\bm x)}{\beta (\bm x)} - 
          \frac{\alpha_i (\bm x)}{\beta (\bm x)} 
          \frac{d x^i}{d \theta} 
     \quad \Rightarrow \quad 
          p_v = 
          \beta (\bm x) 
          \frac{d v}{d \theta} + 
          \alpha_j (\bm x) 
          \frac{d x^j}{d \theta} + 
          U (\bm x), 
     \nonumber 
     \\ 
     \label{jliftmom} 
          \frac{d x^i}{d \theta} &= 
          \frac{\partial \phi}{\partial p_i} = 
          G^{ij} (\bm x) 
          \pi_j 
     \quad \Rightarrow \quad 
          p_i = 
          g_{ij} (\bm x) 
          \frac{d x^j}{d \theta} + 
          \alpha_i (\bm x) 
          \frac{d v}{d \theta} + 
          A_i (\bm x), 
\end{align} 
where 
$g_{ij} (\bm x) = G_{ij} (\bm x) + 
\frac{\alpha_i (\bm x) \alpha_j (\bm x)}{\beta (\bm x)}$
Thus, by applying \eqref{jlift}, \eqref{jliftmom}, and the 
Maupertuis principle \eqref{maup}, I complete the Jacobi 
lift by writing: 
$$
     d s = 
     p_i dx^i + p_v dv = 
     \Omega_{\mu \nu} (\bm x) 
     \frac{d x^\nu}{d \theta} 
     d x^\mu + 
     A_\mu (\bm x) 
     d x^\mu,
$$
\begin{equation} \label{jlift1} 
          d s = 
          \sqrt{
               g_{ij} (\bm x) 
               d x^i d x^j + 
               2 \alpha_i (\bm x) 
               d x^i dv + 
               \beta (\bm x) 
               (d v)^2
          } + 
          A_i (\bm x) 
          d x^i + 
          U (\bm x) 
          d v.
\end{equation}
To lift a RF metric in $n$ co-ordinates to a Riemannian metric 
in $n+1$ co-ordinates, we simply identify the conformal factor 
and gauge fields in \eqref{org} and \eqref{confactor} such that 
$A_i (\bm x) = U (\bm x) = 0$ 
$$
     \frac{\widetilde \beta (\bm x)}{q^2} = 
     \frac{\beta (\bm x)}{(k - U (\bm x))^2} 
\qquad , \qquad 
     q 
     \frac{\widetilde \alpha_i (\bm x)}{\widetilde \beta (\bm x)} = 
     A_i (\bm x) + 
     \frac{\alpha_i (\bm x)}{\beta (\bm x)} 
     \left(k - U (\bm x) \right),
$$ 
\begin{equation} \label{org2} 
     d s = 
     \sqrt{
          \left( 1 - \frac{q^2}{\widetilde \beta (\bm x)} \right) 
          G_{ij} (\bm x) 
          d x^i d x^j
     } + 
     q 
     \frac{\widetilde \alpha_i (\bm x)}{\widetilde \beta (\bm x)} 
     d x^i, 
\end{equation}
such that we get the Riemannian metric:
\begin{equation} \label{jlift2} 
          d s^2 = 
          \left( G_{ij} (\bm x) + 
          \frac{\widetilde \alpha_i (\bm x) \widetilde \alpha_j (\bm x)}
          {\widetilde \beta (\bm x)} \right) 
          d x^i d x^j + 
          2 \widetilde \alpha_i (\bm x) 
          d x^i dv + 
          \widetilde \beta (\bm x) 
          (d v)^2.
\end{equation} 
We will next discuss the nature of RF metrics 
that share the same JMRF.

\subsection{Sharing the JMRF} 
\label{sec:symmrep}

Another alternative to geometrise the potentials 
of the additive term $\bm A (\bm x)$ of the RF 
metric \eqref{rfmetric} is to absorb them into 
$g_{\mu \nu} (\bm x)$, thus converting it into 
a Riemannian metric with the same number 
of co-ordinates. 
However, instead of inserting a new direction 
of symmetry as with the Eisenhart lift, this 
procedure requires identifying a pre-existing 
one and either replacing or rescaling it. 
In effect, we will be describing how to formulate 
all the RF metrics sharing a common JMRF. 
\bigskip 

\noindent 
Let us revisit the Riemannian metric \eqref{riemlag} 
previously discussed: 
$$
     d s^2_R = 
     G_{ij} (\bm x) 
     d x^i 
     d x^j + 
     2 G_{i0} (\bm x) 
     d x^i 
     d T + 
     G_{00} (\bm x) 
     d T^2. 
$$ 
and consider its corresponding Jacobi metric by setting 
$A_\mu (\bm x) = 0$ in \eqref{rfjmet}: 
\begin{equation} \label{riemjmrf} 
     d s_J = 
     \sqrt{
          \left( 
               1 - 
               p_y^2 - 
               \frac{k^2}{G_{00} (\bm x)} 
          \right) 
          \mathbb Y_{ij} (\bm x) d x^i d x^j
     } + 
     k 
     \frac{G_{0i} (\bm x)}{G_{00} (\bm x)} 
     d x^i. 
\end{equation} 
Now let us suppose that the RF metric 
\eqref{randermet} and the Riemannian metric 
\eqref{riemlag} share the same JMRF 
such that \eqref{riemjmrf} matches \eqref{rfjmet}. 
Doing so would also equate the two constraints 
\eqref{constr} and \eqref{riemconstr}, showing that 
a RF metric can be equated to a Riemannian metric 
so long as both have at least one cyclic co-ordinate, 
by writing: 
$$ 
     \sqrt{
          \left( 
               1 - 
               \frac{k^2}{G_{00} (\bm x)} 
          \right) 
          \mathbb Y_{ij} (\bm x) d x^i d x^j
     } + 
     k 
     \frac{G_{0i} (\bm x)}{G_{00} (\bm x)} 
     d x^i = 
     \sqrt{
          \left( 
               1 - 
               \frac{( Q (\bm x) )^2}{g_{00} (\bm x)} 
          \right) 
          \gamma_{ij} (\bm x) d x^i d x^j
     } + 
     \left( 
          A_i (\bm x) + 
          \frac{g_{0i} (\bm x)}{g_{00} (\bm x)} 
          Q (\bm x) 
     \right) 
     d x^i. 
$$ 
\begin{equation} \label{rfelift} 
     \begin{split} 
          G_{00} (\bm x) &= 
          \left( 
               \frac{k}{Q (\bm x)} 
          \right)^2 
          g_{00} (\bm x), 
     \\ 
          G_{i0} (\bm x) &= 
          \frac{k}{Q (\bm x)} 
          \left( 
               g_{i0} (\bm x) + 
               \frac{A_i (\bm x)}{Q (\bm x)} 
               g_{00} (\bm x) 
          \right),
     \\ 
          G_{ij} (\bm x) &= 
          g_{ij} (\bm x) + 
          \frac{A_j (\bm x)}{Q (\bm x)} 
          g_{i0} (\bm x) + 
          \frac{A_i (\bm x)}{Q (\bm x)} 
          g_{j0} (\bm x) + 
          \frac{A_i (\bm x) A_j (\bm x)}{\left(Q (\bm x)\right)^2} 
          g_{00} (\bm x) . 
     \end{split} 
\end{equation}  
So according to \eqref{rfelift} we have the Riemannian 
metric:
\begin{align} 
     d s^2_R &=  
     \left( 
          g_{ij} (\bm x) + 
          \Sigma_{ij} (\bm x) 
     \right) 
     d x^i d x^j 
\nonumber 
\\ 
     \label{eliftact2} 
     & \hspace{0.75cm} + 
     2 \frac{k}{Q (\bm x)} 
     \left(
          g_{i0} (\bm x) + 
     \frac{A_i (\bm x)}{Q (\bm x)} 
     g_{00} (\bm x)
     \right) 
     d x^i d T + 
     \left( 
          \frac{k}{Q (\bm x)} 
     \right)^2 
     g_{00} (\bm x) 
     d T^2, 
\end{align} 
$$
     \text{where } \quad 
     \Sigma_{ij} (\bm x) = 
     \frac{A_j (\bm x)}{Q (\bm x)} 
     g_{i0} (\bm x) + 
     \frac{A_i (\bm x)}{Q (\bm x)} 
     g_{j0} (\bm x) + 
     \frac{A_i (\bm x) A_j (\bm x)}{\left(Q (\bm x)\right)^2} 
     g_{00} (\bm x). 
$$
The shared constraint for the two metrics \eqref{randermet} 
and \eqref{riemlag} is given by: 
\begin{equation} \label{eliftcons} 
     \phi (\bm x, \bm p) = 
     \sqrt{
          g^{ij} (\bm x) 
          \Pi_i \Pi_j + 
          \frac{(Q (\bm x))^2}{g_{00} (\bm x)}
     } = 
     \sqrt{
          G^{ij} (\bm x) 
          \Pi^*_i \Pi^*_j + 
          \frac{k^2}{G_{00} (\bm x)}
     } = 1. 
\end{equation} 
As we can see from \eqref{eliftact2}, the signature 
of the metric is preserved, meaning that if $t$ is time, 
then $T$ can be treated as a rescaled time.
Furthermore, since 
$\psi (\bm x, \bm p, p_T) = \phi (\bm x, \bm p, p_t)$, 
according to the first equation of \eqref{conseq}, 
we can write 
\begin{equation} \label{para} 
     \frac{\partial \phi}{\partial p_i} = 
     \frac{\partial \psi}{\partial p_i} 
\qquad \Rightarrow \qquad 
     \frac{d x^i}{d \sigma} = 
     \frac{d x^i}{d \lambda}
\qquad \Rightarrow \qquad 
     d \sigma = d \lambda.
\end{equation} 
Applying \eqref{maup2} to both, RF \eqref{randermet} 
and Riemannian \eqref{eliftact2} metrics, we can say 
that according to \eqref{para} 
\begin{equation} \label{tscale1} 
     \left. \begin{split} 
          d s_{RF} &= 
          p_i 
          d x^i + 
          q \; 
          d t = 
          d \sigma + 
          A_\mu (\bm x) 
          d x^\mu, 
     \\  
          d s_R &= 
          p_i d x^i + 
          k \; 
          d T = 
          d \lambda, 
     \end{split} \right\} 
\ \Rightarrow \ 
     \frac{d T}{d t} = 
     \omega (\bm x) - 
     \alpha_i (\bm x) 
     \frac{d x^i}{d t}.
\end{equation} 
where 
$$
     Q (\bm x) = k \; \omega (\bm x) 
\qquad , \qquad 
     A_i (\bm x) = k \; \alpha_i (\bm x), 
$$
If $A_i (\bm x) = 0$, then the time rescaling is position dependent 
\begin{equation} \label{tscale2} 
     A_i (\bm x) = 0 
\qquad \Rightarrow \qquad 
     \frac{d T}{d t} = 
     \omega (\bm x).
\end{equation} 
While the metric may not have been lifted by increasing the 
number of canonical pairs, I have converted the action from 
the RF form into the Riemannian action form free of gauge 
fields. 
\bigskip

\noindent
On the other hand, under the circumstances that one deals with 
a static RF metric 
\begin{equation} \label{staticrf} 
     d s_{RF} = 
     \sqrt{
          g_{ij} (\bm x) d x^i d x^j + 
          g_{00} (\bm x) d t^2
     } + 
     A_i (\bm x) d x^i + A_0 (\bm x) d t,
\end{equation}  
this setting \eqref{eliftact2} produces a stationary spactime metric. 
\begin{equation} \label{staticsr} 
          d s^2_R =  
          \left( 
               g_{ij} (\bm x) + 
               \frac{g_{00} (\bm x)}{(Q (\bm x))^2} 
               A_i (\bm x) A_j (\bm x) 
          \right) 
          d x^i d x^j + 
          2 k 
          \frac{g_{00} (\bm x)}{(Q (\bm x))^2} 
          A_i (\bm x)
          d x^i d T + 
          k^2 
          \frac{g_{00} (\bm x)}{(Q (\bm x))^2} 
          d T^2, 
\end{equation} 
thus, supporting the interpretation that the linear terms of a RF metric 
are comparable to the potential terms of a vector potential, and that 
motion in a spacetime described by a stationary metric is comparable 
to motion in the presence of a magnetic field. 
Furthermore, if we set that $q = k$ and $A_0 (\bm x) = 0$, 
then \eqref{staticsr} will become: 
\begin{equation} \label{staticsr2} 
          d s^2_R =  
          \left( 
               g_{ij} (\bm x) + 
               g_{00} (\bm x) 
               \alpha_i (\bm x) \alpha_j (\bm x) 
          \right) 
          d x^i d x^j + 
          2 g_{00} (\bm x) 
          \alpha_i (\bm x)
          d x^i d T_* + 
          g_{00} (\bm x) 
          d T^2_*, 
\end{equation} 
where the time according to \eqref{tscale1} is given by: 
\begin{equation} \label{newtime} 
     \frac{d T_*}{d t} = 
     1 - 
     \alpha_i (\bm x) 
     \frac{d x^i}{d t}.
\end{equation} 
An interesting example to consider is the Schwarzschild metric 
described in Gullstrand-Painlev\'e (GP) co-ordinates 
\cite{painleve,gullstrand,lemaitre}. 
This is an example where a static metric appears in stationary 
form due to co-ordinate transformation of the time, 
and as a result has an apparent magnetic field influencing motion.

\subsubsection{Kerr metric} 
\label{sec:kerr}

Here, I shall briefly discuss the Kerr metric discussed 
previously in \cite{cggmw}. 
The Kerr metric describes a rotating uncharged black hole 
that is a generalisation of the Schwarzschild black hole 
to include rotation, the exact solution of which was 
discovered by Kerr in 1963 \cite{kerr}. 
The Kerr black hole is readily used as a basic example when 
discussing the theory of frame dragging effect that occurs 
around rotating masses. 
\bigskip 

\noindent 
The Kerr metric in Boyer-Lindquist co-ordinates is given by: 
\begin{equation} \label{kerrmetric} 
          d s^2_R = 
          \left( 
               1 - \frac{2 M r}{\rho^2} 
          \right) 
          d T^2 + 
          \frac{4 M a r \sin^2 \theta}{\rho^2} 
          d \varphi \; d T - 
          \rho^2 
          \left[ 
               \frac{dr^2}{\Delta} + d \theta^2 + 
               \frac{\sin^2 \theta}{\rho^4} 
              \left\{ 
                    \left( 
                          r^2 + a^2 
                    \right)^2 - 
                    a^2 \Delta \sin^2 \theta 
               \right\} 
               d \varphi^2 
          \right], 
\end{equation} 
where $\Delta (r) = r^2 - 2 M r + a^2$ , 
$\rho^2 (r, \theta) = r^2 + a^2 \cos^2 \theta$. 
If we compare \eqref{kerrmetric} to the form of \eqref{staticsr2} 
for $q = k, A_0 (\bm x) = 0$, then we shall have: 
$$
     g_{00} (\bm x) = 
     1 - \frac{2 M r}{\rho^2} 
\quad , \quad 
     A_\varphi = 
     k 
     \frac{2 M a r \sin^2 \theta}{\Delta - a^2 \sin^2 \theta} 
$$ 
and its corresponding symmetry replaced RF form 
according to \eqref{staticrf} will be: 
\begin{equation} \label{rfkerr} 
          d s_{RF} = 
          \sqrt{
               \left( 
                    1 - \frac{2 M r}{\rho^2} 
               \right) d t^2 - 
               \rho^2 
               \left[ 
                    \frac{dr^2}{\Delta} + d \theta^2 + 
                    \frac{\Delta \sin^2 \theta}{\Delta - a^2 \sin^2 \theta} 
                    d \varphi^2 
               \right]
          } + 
          k 
          \frac{2 M a r \sin^2 \theta}{\Delta - a^2 \sin^2 \theta} 
          d \varphi, 
\end{equation} 
where the reparametrisation rule \eqref{newtime} 
is written as: 
\begin{equation} \label{repara1} 
     \frac{d T}{d t} = 
     1 - A_\varphi 
     \frac{d \varphi}{d t}. 
\end{equation} 
From the metric \eqref{rfkerr}, 
we can write the reparametrisation formula 
\eqref{repara1} as: 
$$
     p_\varphi = 
     - \rho^2 
     \frac{\Delta \sin^2 \theta}{\Delta - a^2 \sin^2 \theta} 
     \frac{d \varphi}{d \sigma} + 
     k 
     \frac{2 M a r \sin^2 \theta}{\Delta - a^2 \sin^2 \theta} = 
     l 
\qquad , \qquad 
     p_0 = 
     \left( 
          1 - \frac{2 M r}{\rho^2} 
     \right) 
     \frac{d t}{d \sigma} = 
     k
$$ 
$$
     \Rightarrow \qquad 
     \frac{d \varphi}{d t} = 
     - \frac{2 k M a r \sin^2 \theta - 
     l 
     \left( 
          \rho^2 - 2 M r 
     \right)} 
     {k \rho^4 \Delta \sin^2 \theta} 
     \left( 
          \rho^2 - 2 M r 
     \right) 
$$ 
\begin{equation} \label{repara2} 
     \frac{d T}{d t} = 
     1 + 
     2 M a r 
     \frac{2 k M a r \sin^2 \theta - 
     l 
     \left( 
          \rho^2 - 2 M r 
     \right)} 
     {\rho^4 \Delta}. 
\end{equation} 
Thus, we have shown that the Kerr spacetime 
is comparable to a static spacetime with a magnetic 
field generated by a magnetic dipole. 
Naturally, upon setting $a = 0$, we recover 
the Schwarzschild metric.

\subsubsection{Schwarzschild Gullstrand-Painlev\'e metric} 
\label{sec:schwarz.pain} 

\noindent 
This time, we shall consider an interesting example of the 
Schwarzschild metric in Gullstrand-Painlev\'e co-ordinates. 
Let us start by considering the Schwarzschild metric 
in its regular form: 
\begin{equation} \label{schwz} 
     d s^2_R = 
     f (r) 
     d t^2 - 
     \frac1{f (r)} 
     d r^2 - 
     r^2 \left( 
          d \theta^2 + \sin^2 \theta \; d \varphi^2 
     \right) 
, \qquad 
     \text{where } \ 
     f (r) = 1 - \frac{r_0}r, 
\end{equation} 
for which the conserved momentum associated with time $t$ 
according to \eqref{rfgcmom2} is: 
\begin{equation} \label{schwzmom} 
     p_0 = 
     f (r) \frac{d t}{d s} = 
     k (const). 
\end{equation} 
The GP co-ordinate system $T = t - a (r)$, $a (r)$ being some 
function is meant to describe the metric as observed by a radially 
infalling observer. 
\begin{equation} \label{schwzgp} 
     d s^2_R = 
     f (r) 
     \left( 
          d T^2 + 
          2 a' (r) d T d r 
     \right) - 
     \left[ 
          \frac1{f (r)} - 
          f (r) 
          \left( a' (r) \right)^2 
     \right] 
     d r^2 - 
     r^2 \left( 
          d \theta^2 + \sin^2 \theta \; d \varphi^2 
     \right). 
\end{equation} 
The momentum associated with $T$ according 
to \eqref{rfgcmom2} is given by: 
\begin{equation} \label{gpmom} 
     P_0 = 
     f (r) 
     \left( 
          \frac{d T}{d s} + 
          a' (r) \frac{d r}{d s}
     \right) = 
     f (r) \frac{d t}{d s} = 
     k (const). 
\end{equation} 
which according to \eqref{schwzmom} is the same 
value of constant conserved momentum, implying 
that we can apply the theory of Jacobi metric 
sharing to Schwarzschild metric in GP co-ordinates. 
If we compare \eqref{schwzgp} to \eqref{staticsr} 
and choose to set $q = k$ and $A_0 (\bm x) = 0$, 
then we can write \eqref{staticrf} as: 
\begin{equation} \label{schpainrf1} 
     d s_{RF} = 
     \sqrt{
          f (r)  
          d {\widetilde t}^2 - 
          \frac1{f (r)} 
          d r^2 - 
          r^2 
          \left( 
               d \theta^2 + \sin^2 \theta \; d \varphi^2 
          \right)
     } + 
     k 
     a' (r) d r,
\end{equation} 
where we can see that the additive linear term 
at the end of \eqref{schpainrf1} is a gradient 
of the function $a (r)$, and is dismissible 
according to Lagrangian mechanics. 
We can also see that since we have deduced 
a RF metric from the Riemannian Schwarzschild 
GP metric, we have according to \eqref{tscale1}: 
$$
     \frac{d T}{d \widetilde t} = 
     1 - a' (r) \frac{d r}{d \widetilde t}  
\qquad \Rightarrow \qquad 
     \widetilde t = T + a (r) = t. 
$$
thus showing that the Schwarzschild metric 
in GP co-ordinates is essentially no different 
from the default Schwarzschild metric \eqref{schwz} 
with a linear gradient term added. 
If one were to deduce the Jacobi metric for 
the Schwarzschild metric in GP co-ordinates 
\eqref{schwzgp}, then we will have according 
to \eqref{rfjmet}: 
\begin{equation} \label{sgpjmet} 
     d s_J = 
     \sqrt{
          - \left( 
               1 - \frac1{f (r)} 
          \right) 
          \left[ 
          \frac1{f (r)} 
          d r^2 + 
          r^2 \left( 
          d \theta^2 + \sin^2 \theta \; d \varphi^2 
          \right)
          \right]
     } + 
     k 
     a' (r) d r,
\end{equation} 
where again, the linear additive term outside 
square root is a dismissible gradient term, 
showing that the final Jacobi metric is the same 
as that of the familiar Schwarzschild metric.

\section{Conclusion and Discussion} 

I showed that the conventional approach to 
Hamiltonian mechanics via Legendre's principle 
faces a limitation when dealing with RF metrics. 
The solution is to formulate the momentum 
constraint as a suitable and more general 
alternative to the conventional Hamiltonian 
as a generator of Hamilton's equations 
of motion. 
A simple modification of the metric that adds 
an auxiliary co-ordinate makes the constraint 
suitable for studying light-like curves as well. 
This formulation is a simple and significant 
improvement in the generalisation of Hamiltonian 
mechanics when dealing with RF spacetimes 
in relativistic settings. 
With the emerging frequent discussion of RF 
spacetimes, such theories will prove to be quite 
useful tools in their study.

Next, I deduced the generalised Jacobi metric 
known as the JMRF metric for a given RF metric 
for an autonomous relativistic system using the 
constraint. 
For the settings for massive particles, I have 
discussed two familiar settings as examples 
to reproduce previous familiar Jacobi metrics. 
For the settings for null curves, I showed that 
the JMRF metric is distinct from the optical metric 
deduced according to Fermat's principle of path 
of least time for stationary spacetimes. 
This distinction implies that calculating the 
deflection of light rays when studying gravitational 
lensing in stationary spacetimes warrants caution 
since the results will diverge depending on 
whether the optical JMRF or the Fermat metric 
was used. 
Further exploration is required to determine 
which metric is the correct for such calculations.

I have also discussed the frame dragging effect 
from a Hamiltonian mechanics approach using 
Hamiltonian mechanics with the constraint. 
Since frame dragging manifests from the cross 
terms of stationary spacetimes, I derived it for 
RF metrics and the JMRF metric. 
Here, we can see that mechanics with the 
constraint is more suitable than with the 
Hamiltonian since the case of the Jacobi metric 
has no Hamiltonian. 
Given the significance of frame dragging effect 
as an observable and measurable effect predicted 
by general relativity, this discussion opens the 
window for possible application of the Jacobi 
metric into related studies.

Finally, I showed that the Eisenhart lift cannot 
be directly applied to an RF metric in the same manner 
it is for non-relativistic problems. 
Instead, there are 2 alternatives to geometrise 
of the RF metric's linear term potentials. 
The first is to identify the RF metric as a JMRF 
metric and reverse the derivation process to lift 
it into a Riemannian metric, 
thus geometrising the gauge potentials. 
For the second, I discussed autonomous pairs 
of RF and Riemannian metrics that share a common 
JMRF. 
These two new methods of geometrisation present 
useful alternatives to Eisenhart lift when dealing with 
relativistic systems, allowing application of the technique 
beyond the usual non-relativistic setting. 
The case of stationary Riemannian metric and 
a static RF metric shows that the cross terms are 
dynamically comparable to magnetic gauge fields 
as demonstrated with the example of the Kerr metric. 
This suggests another approach to describe and 
study the theory of gravitomagnetism that manifests 
in stationary spacetimes. 
When applied to Schwarzschild Painlev\'e metric 
to derive a static RF metric, we see that the cross 
term introduced via co-ordinate transformation 
is comparable to a total function derivative, 
which is dismissible from any Lagrangian. 

\bigskip

\noindent
\textbf{Acknowledgements }
I wish to acknowledge P. Guha, G. W. Gibbons, 
P. Maraner, M. Werner, M. Cariglia and Sanved 
Kolekar for various discussions related to this 
topic and their support that was essential in the 
preparation of this article, and thank P. Horvathy, 
K. Morand, A. Galajinsky, E. Minguzzi, M.F. Ranada, 
Joydeep Chakravarty, and the anonymous reviewer 
for supportive comments that helped improve and 
develop its content.

%
% For  figures use
%\begin{figure*}
%\vspace*{5cm}       % Give the correct figure height in cm
%\includegraphics{leer.eps}
%\end{figure*}
% or  this
%\begin{figure}
%\centering
%\resizebox{0.75\textwidth}{!}{%
%  \includegraphics{leer.eps}
%\vspace{5cm}       % Give the correct figure height in cm
%\end{figure}
%\begin{table}
%\centering
%\begin{tabular}{lll}
%\hline\noalign{\smallskip}
%\end{tabular}
% Or use
%\vspace*{5cm}  % with the correct table height
%\end{table}

%
% BibTeX users please use
% \bibliographystyle{}
% \bibliography{}

\end{document}